\documentclass[%
aps,
pre,
superscriptaddress,
tightenlines,
showpacs,showkeys,
a4paper,
12pt,
longbibliography,
notitlepage
]{revtex4-1}
\usepackage[english]{babel}
\usepackage{dcolumn}
\usepackage{amssymb,amsmath,stmaryrd,bm}

\usepackage{graphicx}
\usepackage{subfig}

\makeatletter

\newcommand{\ket}[1]{\ensuremath{\bigl|{#1}\bigr\rangle}}
\newcommand{\bra}[1]{\ensuremath{\bigl\langle{#1}\bigr|}}
\newcommand{\qsca}[2]{\ensuremath{\bigl\langle{#1}|{#2}\bigr\rangle}}

\newcommand{\avr}[1]{\ensuremath{\bigl\langle{#1}\bigr\rangle}}

\newcommand{\oprt}[1]{\ensuremath{\widehat{\mathcal{#1}}}}

\newcommand{\hcnj}[1]{{#1}^{\dagger}}

\newcommand{\pdrs}[1]{\partial_{#1}}
\newcommand{\pdr}[2]{\dfrac{\partial #1}{\partial #2}}

 \newcommand{\vc}[1]{\mathbf{#1}}

\newcommand{\dd}{\mathrm{d}}
\newcommand{\DD}{\ensuremath{\mathcal{D}}}
 
 \newcommand{\ee}{\mathrm{e}}

\newcommand{\eff}{\mathrm{eff}}

\newcommand{\st}{\mathrm{st}}
\newcommand{\eqlb}{\mathrm{eq}}

\newcommand{\med}{\mathrm{m}}

\makeatother

\begin{document}
\title{Energy representation  for  nonequilibrium Brownian-like
  systems: 
steady states and fluctuation relations}

\author{Bohdan~I.~Lev}
\email[Email address: ]{bohdan.lev@gmail.com}

\affiliation{M.M. Bogolyubov Institute for Theoretical Physics of the NAS
of Ukraine, 14-b, Metrolohichna Str., Ky\"{\i}v, Ukraine}

\author{Alexei~D.~Kiselev}
\email[Email address: ]{kiselev@iop.kiev.ua}

\affiliation{%
 Institute of Physics of National Academy of Sciences of Ukraine,
 prospekt Nauki 46,
 03028 Ky\"{\i}v, Ukraine}

\affiliation{M.M. Bogolyubov Institute for Theoretical Physics of the NAS
of Ukraine, 14-b, Metrolohichna Str., Ky\"{\i}v, Ukraine}

\date{\today}
%\date{October 5, 2009} 


\begin{abstract}
Stochastic dynamics in the energy representation
is  used as a novel method to
represent nonequilibrium Brownian-like systems.
It is shown that the equation of motion for the energy of such systems
can be taken in the form of the Langevin equation with multiplicative noise.
Properties of the steady states are examined by solving
the Fokker-Planck equation for the energy distribution functions.
The generalized integral fluctuation theorem
is deduced for the systems
characterized by the shifted probability flux operator.
From this theorem, a number of entropy and fluctuation relations 
such as the Evans-Searles fluctuation theorem, 
the Hatano-Sasa identity and the Jarzynski's equality
are derived.
\end{abstract}

\maketitle
  
%%%%%%%%%%%%%%%%
\section{Introduction}
\label{sec:intro}
%%%%%%%%%%%%%%%%

According to the basic principles of thermodynamics, 
when a macroscopic system is brought into contact with a thermostat
(reservoir, heat bath), the system evolves in time approaching
the equilibrium state in the course of relaxation.  
The state of equilibrium is well-defined only
under certain idealized  
conditions~\cite{Huang:bk:1987,Huang:bk:2005,Balescu:bk:1975}
so that, at equilibrium,  
the thermodynamical parameters 
of the system are adjusted to the values of the thermostat
whereas all intrinsic flows have come to an end. 

In most cases, however, systems are subject to
nonequilibrium conditions and external 
constraints~\cite{Groot:bk:1974,Balescu:bk:1975,Balescu:bk:1997,
Mazenko:bk:2006,Zwanzig:bk:2001,Streater:bk:1995,Ross:bk:2008}.
In open systems out of equilibrium, 
flows universally present and generally cannot cease to exist.   
So, it is difficult if not impossible to determine governing 
parameters that can be held constant. 

Nevertheless, there are the stationary states 
that can be unambiguously defined for certain open
systems~\cite{Groot:bk:1974}.
Although such states are independent of time and thus
might be regarded as ``equilibrium'' ones, the
thermodynamical parameters of the system 
may significantly differ from those for the environment 
and, more generally, the stationary distribution functions 
cannot be described using the well-known equilibrium distributions.

Among all the nonequilibrium systems,
the most studied and important case
is represented by an ensemble of Brownian particles. 
In spite of the fact that Brownian motion 
has long been the subject of intense studies
(a recent review on its history can be found in~\cite{Nelson:bk:2001,Mazo:bk:2002}),
it is still interesting to understand the behavior of 
Brownian particles 
as a model system driven far from equilibrium.

It should be emphasized that
the theory of Brownian motion
can be applied to the systems
where the term ``Brownian particle''
does not mean a real particle.
For example, it may point to
some collective property of macroscopic systems
such as the concentration of any component of 
a chemically reacting system~\cite{Zwanzig:bk:2001}. 
For brevity, such systems will be referred to as 
the \textit{Brownian systems}.

Though 
far-from-equilibrium Brownian systems 
are abundant in nature,
there is no unified commonly accepted theoretical approach to
determine possible states of 
such systems.
Hence it is a  fundamentally important task to develop
a method to explore general properties of
stationary states of open systems
and to establish  the conditions of their existence.

In equilibrium statistical mechanics,
these states are known to be generally described 
in terms of the energy surfaces
giving, for certain systems,
microcanonical and canonical ensembles~\cite{Huang:bk:1987,Balescu:bk:1975}. 
In a similar spirit,
our considerations will be based on
the \textit{energy representation} where
the states of the Brownian system are determined 
solely by their energies.

Note that similar representation
has been previously used
in energy controlled stochastic models
such as the random energy model~\cite{Derrida:prl:1980,Derrida:prb:1981}
and its generalizations~\cite{Derrida:jphq:1985}. 
The energy master equation was also 
derived in Ref.~\cite{Dyre:prb:1995}
as an low-temperature approximation describing energy
fluctuations in the B\"{a}ssler's phenomenological
random-walk model for viscous liquids. 

Typically,
the interaction between
the Brownian system such as a Brownian particle
and the environment
involves the process of
direct energy interchange.
During this process,
the dissipation may take the energy away from the system
leading to a loss of its  energy (the positive friction).
In the opposite case of negative friction,  
the transfer of energy from the thermostat
results in the energy input.

In  addition to the deterministic part of interaction,
there are fluctuation effects of the environment
that affect the system giving rise to rapid change of its state. 
Such changes may take place when the structure of the environment
is complicated by the presence of additional systems or
some of its characteristics can be directly influenced by
the processes running in the Brownian system. 
%In a sense, by contrast to a heat reservoir, the environment 
%generally cannot be regarded as a fixed formation.

So, the above pattern suggests
using stochastic dynamics to describe 
the behavior of Brownian systems interacting with
surrounding media.
Therefore, there is
an ensemble of such systems characterized
by the probability distribution function
that, in particular, define stationary (steady) states
formed  under nonequilibrium conditions.
Typically, in systems far from equilibrium, 
the distribution functions
of the stationary states significantly differ from
the well-known equilibrium distributions. 
For example, dissipative dynamics of  
the noise driven Hookean model
for protein folding studied in Ref.~\cite{Tuzel:jsp:2000}
was found to be characterized by 
a distribution of energy states obeying a modified one-dimensional 
Ornstein-Uhlenbeck process.

In this paper we 
suggest using the Langevin dynamics~\cite{Snook:bk:2007,Coffey:bk:2004}
in the energy representation   
to describe the class of systems called
the Brownian systems.
Properties of  the stationary (steady) states of such systems will be of 
our primary concern. 
The layout of the paper is as follows.

In Sec.~\ref{sec:brownian}, we start with 
the purely dissipative dynamics of  Brownian particles
and show that, in the energy representation, 
it is governed by the Langevin equation with 
the multiplicative white noise.
Then we generalize the results for the Brownian particles
and formulate the Langevin equation~\eqref{eq:Langv_energy}
for out-of-equilibrium Brownian-like  systems.

In Sec.~\ref{sec:steady_states},
the  steady states of the Brownian systems
are obtained as the stationary solutions of the Fokker-Planck
equation and are characterized by the probability flux number, $J_{\st}$.
It is found that the steady state distributions are
additionally determined by
the effective energy
potential~\eqref{eq:V_potential} 
and 
by the energy dependent diffusion coefficient.
We also discuss some important
examples of the steady states
formed when the probability flux
vanishes, $J_{\st}=0$, and the
distributions take the potential form.

In Sec.~\ref{subsec:entropy},
we  define the trajectory
dependent entropies
whose ensemble averages can be
associated with the entropy
production rates of the system and 
of the environment.
The steady state values of the rates are evaluated. 
The rate of the medium entropy 
is found to depend heavily on the
effective potential.
Then, 
in  Sec.~\ref{subsec:fluc-rels},
we derive 
the generalized integral fluctuation 
relation~\eqref{eq:gen-HS-idnt} 
for Brownian systems 
with the shifted probability flux operator 
characterized by the flux parameter and 
the probability flux number.
It is shown that 
there are a number of 
the known fluctuation relations 
along with the fluctuation-dissipation
theorem
for the steady-state systems
that immediately follow from 
the relation~\eqref{eq:gen-HS-idnt}.  

We discuss our results and make some concluding remarks
in Sec.~\ref{sec:discuss}.

%%%%%%%%%%%%%%%%%%%%%%%%%%%%%%%%%%%%
\section{Brownian particles and energy representation}
\label{sec:brownian}
%%%%%%%%%%%%%%%%%%%%%%%%%%%%%%%%%%%%

%%%%%%%%%%%%%%%%%%%%%%%%%
\subsection{Brownian particles}
\label{subsec:brwn-particles}
%%%%%%%%%%%%%%%%%%%%%%%%

The energy representation, 
where the states of the system are 
distinguished solely by their energies, $\varepsilon$,
can be viewed as
the very basic description of 
the nonequilibrium system behavior provided that
other dynamical variables, for some reasons, 
are irrelevant and can be disregarded.
In this section, 
the well-known case of 
purely dissipative dynamics of a Brownian
particle~\cite{Just:bk4:2002}
will be our initial concern.
For this simple model system,
we introduce the energy representation
by deriving the Langevin equation 
 for the energy of the particle.

So, we take, as the starting point,
the following Langevin equation  
  \begin{equation}
\label{eq:Langv_brownian_3d}
\pdrs{t} p_i=F_i+\eta_i,
\end{equation}
where  $\vc{p}=(p_1,p_2,p_3)$ is the momentum of the particle;
$F_i= -\pdr{E}{p_i}\equiv -\pdrs{i} E$ is the $i$th component of
 the force expressed in terms of the potential, $E\equiv E(p)$.
Note that the linear friction force $\vc{F}=-\gamma \vc{p}$ represents 
the special case where the potential is proportional to
the kinetic energy, $\varepsilon=p^2/(2M)$ 
($M$ is the particle mass): $E=\gamma p^2/2 = \gamma M  \varepsilon$,
($\gamma$ is the friction coefficient). 

For the white Gaussian noise with $\avr{\eta_{i}(t)}=0$
and $\avr{\eta_i(t)\eta_j(t')}=2 \sigma^2\delta_{ij}\delta(t-t')$,
where $2 \sigma^2$ is the intensity of the random force,
Eq.~\eqref{eq:Langv_brownian_3d} is known to give
the Fokker-Planck (FP) equation
\begin{align}
  \label{eq:FP_brownian_3d}
 \pdrs{t} P(\vc{p},t)=\pdrs{i} 
\Bigl[
\sigma^2\,\pdrs{i} - F_i
\Bigr]\,P(\vc{p},t)
\end{align}
describing the stochastic dynamics as the evolution
of the probability distribution function
$P(\vc{p},t)=\avr{\delta(\vc{p}(t)-\vc{p})}$.
The equilibrium distribution
\begin{align}
  \label{eq:P_eq_brownian_3d}
  P_{\eqlb}(\vc{p})\propto\exp[- E(p)/\sigma^{2}]
\end{align}
then can be derived as 
the stationary solution to the FP equation~\eqref{eq:FP_brownian_3d}.
For $E=\gamma M \varepsilon$, 
Eq.~\eqref{eq:P_eq_brownian_3d} gives 
the well-known Boltzmann distribution provided
the friction coefficient, $\gamma$, the mass, $M$, and the noise
intensity, $\sigma^2$,
are related to the inverse temperature, $\beta=1/(k_B T)$,
through the Einstein relation: $\gamma M/\sigma^2 = \beta$.

Our task now is to deduce the Langevin equation for 
the energy of the Brownian particle, $\varepsilon$.
To this end we start from the FP equation~\eqref{eq:FP_energy_brownian}
written in the spherical coordinates $(p,\theta,\phi)$.
Then, after averaging over angles and making the change of
variables, $p\to \varepsilon=p^2/(2M)$, we have 
\begin{align}
  \label{eq:FP_energy_brownian}
  \pdrs{t} \tilde{P}(\varepsilon,t)=\pdrs{\varepsilon} 
\Bigl[
\tilde{\sigma}^2\,\sqrt{\varepsilon}\,\pdrs{\varepsilon}\,\sqrt{\varepsilon} 
+ (\varepsilon \tilde{E}^{\,\prime}-\tilde{\sigma}^2)
\Bigr]\,\tilde{P}(\varepsilon,t),
\end{align}
where $\tilde{\sigma}^2=\sigma^2/(8M)$,
$\tilde{E}=E/(8M)$, 
$\tilde{P}(\varepsilon,t)$ is the energy distribution function
and prime stands for the derivative with respect to
the energy, $\varepsilon$. 

The steady state distribution 
found as the stationary solution of 
the FP equation in the energy
representation~\eqref{eq:FP_energy_brownian} 
is given by
\begin{align}
  \label{eq:Peq_energy_brownian}
  \tilde{P}_{\eqlb}(\varepsilon)\propto\sqrt{\varepsilon}\,
\exp[-\tilde{\sigma}^{-2}\tilde{E}(\varepsilon)].
\end{align}
The difference between the distributions given in
Eq.~\eqref{eq:P_eq_brownian_3d}
and Eq.~\eqref{eq:Peq_energy_brownian}
is due to the additional square root factor $\sqrt{\varepsilon}$
that accounts for the Jacobian of the transformation: 
$p^2\dd p\propto \sqrt{\varepsilon}\,\dd\varepsilon$.

For the Boltzmann equilibrium distribution
with $E/\sigma^2=\tilde{E}/\tilde{\sigma}^2=\beta \varepsilon$,
it can be readily seen that
the function~\eqref{eq:Peq_energy_brownian} 
reaches its maximum value at $\varepsilon=k_{B}T/2$
and the equilibrium mean value of the energy 
is $\avr{\varepsilon}_{\eqlb}= 3 k_{B} T/2$. 
This case represents the equilibrium conditions, 
when the Brownian system is at equilibrium and 
the environment plays the role of the thermostat. 

Finally, from Eq.~\eqref{eq:FP_energy_brownian},
it can be inferred that the Langevin equation
in the energy representation is given by
\begin{align}
  \label{eq:Lang_energy_brownian}
  \pdrs{t}\varepsilon = - (\varepsilon
  \tilde{E}^{\,\prime}-\tilde{\sigma}^2)
+\sqrt{\varepsilon}\,\tilde{\eta}(t),
\quad
\avr{\tilde{\eta}(t)\tilde{\eta}(t')}= 2\tilde{\sigma}^2\,\delta(t-t').
\end{align}
This result shows that the stochastic 
equation governing the dynamics of the Brownian particle energy is 
characterized by the multiplicative noise.

Note that, though different representations of the Brownian motion are equivalent, 
the energy representation 
can be the preferential approach 
when it is necessary to take into account
both loss and gain (dissipation and absorption)
of the energy.
The latter process implies that the Brownian particle
is subjected to  ``negative'' friction.

We can now draw some generalizations from the results
for the Brownian particle and
treat the general case of Brownian systems.

%%%%%%%%%%%%%%%%%%%%%%%%%%
\subsection{Langevin dynamics in energy representation}
\label{subsec:energy-reprsnt}
%%%%%%%%%%%%%%%%%%%%%%%%%%

Our basic assumption is that,
similar to the above discussed Brownian particle, 
the energy dynamics of the Brownian system
is governed by the Langevin equation with multiplicative noise
of the general form
\begin{equation}
\label{eq:Langv_energy}
\pdrs{t}{\varepsilon}=-f(\varepsilon)+g(\varepsilon)
\xi(t),
\end{equation}
where $\varepsilon$ is the energy, 
$f(\varepsilon)$ is
the function giving
the \textit{rate of direct energy exchange}
and $g(\varepsilon)$ is 
the \textit{energy diffusion function};
$\xi(t)$ represents Gaussian white noise.
Mathematically, this equation can be regarded as the generalized version of 
Eq.~\eqref{eq:Lang_energy_brownian},
where $f(\varepsilon)=\varepsilon \tilde{E}^{\,\prime}-\tilde{\sigma}^2$
and
$g(\varepsilon)=\sqrt{\varepsilon}$. 
 
The first term on the right hand side of 
dynamical equation~\eqref{eq:Langv_energy}
is due to the direct action of the environment on the system,  
The effect of direct external action 
described by the \textit{exchange function}, $f(\varepsilon)$,
is determined by
the conditions under which the system is kept and 
by its physical characteristics.

But evolution of the system state is not
determined solely by these factors. 
Each parameter of the
system may undergo irregular variations
caused either by fluctuation induced stochastic perturbations 
or by the complicated dynamical behavior of a nonlinear environment. 
These additional variations result in random migration of the system 
over various states.

Random influence of the environment
is represented by the second term on the right hand side of
Eq.~\eqref{eq:Langv_energy}.
This term accounts for the system--environment
interaction induced by fluctuations in parameters
of the system controlling conditions.
It is taken in the form of multiplicative noise
with the \textit{diffusion function}, $g(\varepsilon)$,
giving the energy dependent coupling strength. 

Thus,
for  the Brownian system in contact with the environment,
there are processes leading  to gain and loss of the energy 
 that underlie the Langevin dynamics in the energy representations.  
 Our next problem is to identify
the conditions for the system to be at equilibrium stationary states 
and to find the distribution function of the system out-of-equilibrium. 

%%%%%%%%%%%%%%%%%%%%%%%%%%%%%%%%
\section{Nonequilibrium steady states}
\label{sec:steady_states}
%%%%%%%%%%%%%%%%%%%%%%%%%%%%%%%%

In the previous section we have found that,
for the energy of the Brownian particle expressed in terms of 
the momenta, the stochastic dynamics is governed by 
the Langevin equation~\eqref{eq:Lang_energy_brownian}.

In more general nonequilibrium systems,
it does not always happen that all necessary details on
the dynamics of internal degrees of freedom such as 
the momenta are known. 
These  nonequilibrium systems can be characterized
by the energy dependent probability distribution function 
$\rho(\varepsilon,t)=\avr{\delta(\varepsilon(t)-\varepsilon)}$ 
describing the dynamical behavior
that involve the processes of energy loss and gain  induced by 
the environment.

Generally, 
the equation of motion for the energy change
of out-of-equilibrium systems 
is complicated by nonlinearities
present in both  the internal dynamics of the system
and in the system-environment coupling.
For Brownian systems in the energy representation, 
this complexity  can be described using the nonlinear Langevin equation
with the multiplicative noise~\eqref{eq:Langv_energy}
that define the nonlinear stochastic out-of-equilibrium dynamics.

In this section,
we  begin with the dynamics of 
the probability distribution function, $\rho(\varepsilon,t)$,
and  then examine the properties of the steady states.
These are defined as the stationary solutions of 
the FP equation in the energy representation.
This approach to the steady states closely resembles
the widely accepted definition
of the equilibrium distribution function
which implies that, for a typical system,  
the distribution is determined solely by the energy  as 
the only known integral of motion. 

%%%%%%%%%%%%%%%%%%%%%%%%%%
\subsection{Fokker-Planck dynamics}
\label{subsec:fokk-planck-dynam}
%%%%%%%%%%%%%%%%%%%%%%%%%%

 Our first step is to deduce the FP equation
that define how the energy distribution $\rho(\varepsilon,t)$
evolves in time. 
To this end, we 
adopt the symmetric Stratonovich convention and
apply the standard procedure~\cite{Gard,Just:bk4:2002}
to the Langevin equation~\eqref{eq:Langv_energy}
giving the following result:
\begin{equation}
\label{eq:FP_energy}
  \pdrs{t} \rho(\varepsilon,t)=\pdrs{\varepsilon} 
\Bigl[
\sigma^2\,g(\varepsilon)\,\pdrs{\varepsilon}\,g(\varepsilon) 
+ f(\varepsilon)
\Bigr]\,\rho(\varepsilon,t).
\end{equation}
Note that  
there are different interpretations of 
the Langevin equation~\eqref{eq:Langv_energy}
that all boil down to providing the discretization rules
employed to define the stochastic integral:
$
\int_t^{t+\Delta t}g(\varepsilon(\tau))\xi(\tau)\dd\tau.
$
Mathematically rigorous considerations of stochastic differential
equations are mostly based on the It\^{o} calculus
(the prepoint discretization rule),
whereas 
the Stratonovich interpretation (the midpoint discretization rule)
has simpler transformation
properties under a change of variables
and arises naturally when the delta-correlated noise
is treated as the limiting case of a real noise with finite
correlation time (colored noise).
Since the drift term in the FP equation
depends on the prescription for evaluating 
multiplicative noise
(the results for the generalized Stratonovich prescription 
can be found in Ref.~\cite{Lau:pre:2007}),
the problem known as the It\^{o}-Stratonovich dillema 
arises~\cite{Gard,Lefever:bk:1984,Kampen:bk:1992}. 
Mathematically, 
the results for the It\^{o}
and Stratonovich stochastic differential equations
are in one-to-one correspondence.
But this correspondence is system dependent. So,
additional information about the microscopic structure of the
environment is required in order to decide which discretization rule
is physically adequate
(discussion of the dilemma in the context of ``internal and external''
noise is given in Chap.IX.5 of van Kampen's textbook~\cite{Kampen:bk:1992}).

The FP equation
in the energy representation~\eqref{eq:FP_energy} 
can now be conveniently recast into 
the divergence form giving the conservation law
\begin{equation}
\label{eq:FP_flux}
\frac{\partial \rho(\varepsilon,t)}{\partial t}=\frac{\partial
J(\varepsilon,\rho)}{\partial \varepsilon},
\end{equation}
characterized by the probability flux $J$
\begin{align}
&
\label{eq:flux}
J=
D(\varepsilon)
\Bigl\{
V^{\,\prime}(\varepsilon)
+ \pdrs{\varepsilon}
\Bigr\}
\rho\equiv\oprt{J}\cdot\rho
\end{align}
with the flux operator
\begin{align}
  \label{eq:flux_operator}
  \oprt{J}=D(\varepsilon)
\Bigl\{
V^{\,\prime}(\varepsilon)
+ \pdrs{\varepsilon}
\Bigr\},
\end{align}
where
$D(\varepsilon) = \sigma^{2} g^{2}(\varepsilon)$
is the \textit{diffusion coefficient}
and $V$ is the \textit{effective energy potential}
given by
\begin{align}
&
\label{eq:V_potential}
V(\varepsilon)=\ln |g(\varepsilon)|+
\int^{\varepsilon}_{\varepsilon_{0}}
f(\varepsilon')/D(\varepsilon')\dd\varepsilon'.
\end{align}

For the values of energy $\varepsilon$ ranged between $\varepsilon_{\min}$
and $\varepsilon_{\max}$, $\varepsilon\in
[\varepsilon_{\min},\varepsilon_{\max}]$,
the conservation law~\eqref{eq:FP_flux} combined with the normalization condition,
$\int_{\varepsilon_{\min}}^{\varepsilon_{\max}}\rho\,\dd\varepsilon=1$,
gives the condition of conservation for the flow of probability:
$J|_{\varepsilon=\varepsilon_{\min}}=J|_{\varepsilon=\varepsilon_{\max}}\equiv
J_b$.

Temporal evolution of the probability distribution function $\rho$
%that enters Eq.~\eqref{eq:entropy-def}
is governed by the evolution operator, $\oprt{U}(t,t_0)$, 
of the Fokker-Planck equation~\eqref{eq:FP_flux}.
This operator can be found as the solution of 
the following initial value problem:
\begin{align}
&
   \label{eq:FP-oper-evol}
   -\pdrs{t}\, \oprt{U}(t,t_0)=
\oprt{H}\cdot
\oprt{U}(t,t_0),
\quad
\oprt{U}(t_0,t_0)=\oprt{I},
\\
&
\label{eq:FP-Hamilt-flux}
\oprt{H}=
-\pdrs{\varepsilon}\cdot\oprt{J},
\quad
\oprt{J}= D\, [V'+\pdrs{\varepsilon}],  
\end{align}
where $\oprt{I}$ is the identity operator; 
$\oprt{H}$ is the Fokker-Planck operator
that plays the role of the effective Hamiltonian 
related to the probability  flux operator~\eqref{eq:flux_operator}.

Thus the probability distribution $\rho(\varepsilon,t)$
evolves in time under the action of the evolution operator.
This can be conveniently expressed using
the quantum mechanical bracket 
notations as follows~\cite{Just:bk4:2002,Kurchan:jpa:1998,Kurchan:jsm:2007}
 \begin{align}
&
   \label{eq:FP-bra-ket}
   \ket{\rho(t)}=
\oprt{U}(t,t_0)\,\ket{\rho(t_0)},
\quad
\rho(\varepsilon,t)=
\qsca{\varepsilon}{\rho(t)}.
 \end{align}

%%%%%%%%%%%%%%%%%%%%%%%%%%%%
\subsection{Stationary distributions}
\label{subsec:stat-distr}
%%%%%%%%%%%%%%%%%%%%%%%%%%%

The general stationary solution of the FP
equation~\eqref{eq:FP_energy}
\begin{align}
&
  \label{eq:steady_state_gen}
 \rho_{\st}(\varepsilon)=\exp\bigl[-V(\varepsilon)\bigr]
\Bigl\{
N_{\st} + 
J_{\st}
\int_{\varepsilon_0}^{\varepsilon}
\exp\bigl[V(\varepsilon')\bigr]/D(\varepsilon')\dd\varepsilon'
\Bigr\}, 
\end{align}
where 
$J_{\st} =J_{b}(\rho_{\st})$ is the stationary probability current and
$N_{\st}$ is the nomarlization constant,
can be easily obtained by solving 
the first-order linear differential equation:
$\oprt{J}\cdot\rho_{\st}=J_{\st}$.
Note that the bracket form of
the equation for the steady states is 
 \begin{align}
   \label{eq:Jst-braket}
   \oprt{J}\ket{\rho_{\st}}=\ket{J_{\st}},
 \end{align}
where 
$\rho_{\st}(\varepsilon)=
\qsca{\varepsilon}{\rho_{\st}} $ and
$\qsca{\varepsilon}{J_{\st}}=J_{\st}$.

%In what follows 
In the remaining part of the section
we assume reflecting boundary conditions
and restrict ourselves to the important special case
where the stationary flow is absent, $J_{\st}=0$. 
For instance, such boundary conditions apply to the case when the energy
spectrum is unbounded from above, $\varepsilon_{\max}\to\infty$
and $\varepsilon_{\min}\le\varepsilon <\infty$,
and the steady state distribution function rapidly decays to zero with 
the energy: 
$\rho_{\st}\to 0$ at $\varepsilon\to\infty$.

So, the steady state distribution~\eqref{eq:steady_state_gen}
takes the potential form
\begin{equation}
\label{eq:station_sol}
\rho_{\st}(\varepsilon)=Z_{\st}^{-1}\exp\left\{-V(\varepsilon)\right\},
\quad
Z_{\st}= \int_{\varepsilon_{\min}}^{\varepsilon_{\max}} \exp[-V(\varepsilon)]\dd\varepsilon.
\end{equation}
Note that, by analogy with equilibrium systems, 
the quantity $F_{\st}=-\ln Z_{\st}$ sometimes is referred to as 
the effective f ree energy~\cite{Klimont:ufn:1994}.

The shape of the distribution~\eqref{eq:station_sol} is determined by
the effective energy potential
given in Eq.~\eqref{eq:V_potential}.
In particular,
the distribution function reaches its extremal value 
at energies determined by the stationary points,
$\varepsilon=\varepsilon_s$,
of the potential $V$.
These points can be found from the stationarity equation
\begin{equation}
\label{eq:extrem}
V'(\varepsilon_s)=\frac{1}{ 2 D(\varepsilon_{s})}
\left(
D'(\varepsilon_s)+ 2 f(\varepsilon_{s})
\right)=0
\end{equation}
that can be regarded as 
the condition of diffusion-drift balance,
$\sigma^{2} g(\varepsilon_s) g'(\varepsilon_s)=-f(\varepsilon_s)$, 
between the
diffusion over states of the environment and 
the dissipation in the system.
This balance condition gives
the  value of most probable steady state energy,
$\varepsilon=\varepsilon_m$,
which corresponds to the minimum of
the energy potential $V$.

In the vicinity of the most probable energy, 
the steady state distribution can be approximated by 
the Gaussian function  
\begin{equation}
\label{eq:st_sol_gaussian}
\rho_{\st}(\varepsilon)\approx\rho_{G}(\varepsilon) \propto 
%\exp\left\{-U(\varepsilon_{s})\right\}
\rho_{\st}(\varepsilon_{m})
\exp\left[-V''(\varepsilon_{m})(\varepsilon-\varepsilon_{m})^{2}/2\right],
\end{equation}
where 
$V''(\varepsilon_m)=
[
D''(\varepsilon_m)+ 2 f'(\varepsilon_{m})
]
/(2 D(\varepsilon_{m}))
$ 
is the second derivative of the potential with respect to the energy. 

There are a variety of 
typical cases representing 
newly formed steady states of nonequilibrium systems
depending on the dissipation and diffusion functions,
$f(\varepsilon)$ and $g(\varepsilon)$.
Below we discuss some of the most important ones.

We begin with the noiseless case 
by assuming the singular limit  of vanishing diffusion,  $g\to 0$.
Then  temporal evolution of 
the energy distribution function initially prepared at $\varepsilon=\varepsilon_{0}$
with $\rho(\varepsilon,0)=\delta(\varepsilon-\varepsilon_{0})$,
is as follows
\begin{align}
  \label{eq:determ_limit}
  \rho(\varepsilon,t)=\delta(\varepsilon-\varepsilon(\varepsilon_{0},t)),
\end{align}
 where $\varepsilon(\varepsilon_{0},t)$ is the solution of the initial
 value problem: 
 \begin{align}
   \label{eq:Cauchy_prb}
   \pdrs{t}\varepsilon=-f(\varepsilon)\equiv -E^{\,\prime}(\varepsilon),
\quad
\varepsilon(0)=\varepsilon_0.
 \end{align}
Suppose that
there is  a local minimum of
the potential $E(\varepsilon)$
located at $\varepsilon=\varepsilon_s$.
Then the energy $\varepsilon_s$
is the attracting stationary (equilibrium)
point that defines the stationary distribution
$\rho_{\st}(\varepsilon)=\delta(\varepsilon-\varepsilon_s)$.
This implies that,
when the initial value of the energy, $\varepsilon_0$,
falls within the corresponding basin of attraction,
the distribution functions~\eqref{eq:determ_limit}
evolve in time approaching the steady state:
$\rho(\varepsilon,t)\to \delta(\varepsilon-\varepsilon_s)$
at $t\to\infty$.

Interestingly, when the diffusion function is a nonzero constant,
$g(\varepsilon)=g_0\ne 0$,
the steady state distribution is 
$\rho_{\st}(\varepsilon)=N
\exp[-E(\varepsilon)/D_0]$,
where $D_0=\sigma^2g_0^2$,
so that its maxima correspond to the equilibria 
of the potential $E$.
By contrast to the noiseless case, at $g_0\ne 0$,
we can have the steady state even without equilibria.
An important example is 
 the canonical equilibrium  Boltzmann-Gibbs (BG) distribution with
$E(\varepsilon)/D_0=\beta\varepsilon$.

Note that
the steady state function
takes the form of "microcanonical distribution",
$\rho_{\st}(\varepsilon)=\delta(\varepsilon-\varepsilon_0)$,
parametrized by the energy value $\varepsilon_{0}$,
only if the energy is conserved, $f(\varepsilon)=0$, 
and diffusion is absent, $g(\varepsilon)=0$.

The limiting case 
with vanishing exchange function,
$f(\varepsilon)=0$,
describes the diffusion controlled 
systems. 
From Eq.~\eqref{eq:station_sol},
we obtain the stationary distribution function
$\rho_{\st}(\varepsilon)=N |g(\varepsilon)|^{-1}$
expressed in terms of the diffusion function
$g(\varepsilon)$.
The BG distribution,
$\rho_{BG}(\varepsilon)=N \exp[-\beta \varepsilon]$, 
can be realized as the steady state distribution
only if  the diffusion coefficient depends exponentially
on the energy: 
$D(\varepsilon)\propto \exp[ 2 \beta \varepsilon]$, 
where $\beta$ is the inverse temperature. 
The exponential dependence
may emerge as the special feature of
the interaction between the system and the
environment.

The above considerations are also applicable to 
the systems of Brownian particles in a randomly
inhomogeneous environment. 
In such environment, 
some characteristics such as 
the coupling constants and the friction coefficient
may contain stochasticity induced 
contributions
and thus become random variables. 
Examples include 
large particles in the inhomogeneous environment,
impurity particles placed into the dusty plasma, 
as well as the systems whose kinetic properties 
depend nonlinearly on the velocity or the
energy of particles. 

As another simple example, we 
consider what happen when
the second viscosity in mixtures and acoustic flow
comes into play
due to the dependence of the friction
coefficient on the velocity~\cite{Landau:6v:en:1987}. 
In this case we can take the assumption that,
in the Langevin equation~\eqref{eq:Langv_energy} with
the linear exchange function 
$f(\varepsilon)=\gamma\varepsilon$, 
the friction coefficient $\gamma=\avr{\gamma}+\xi(t)$  
is a sum of its average value,
$\avr{\gamma}$, and the noise term $\xi$ 
that accounts for random variations of the friction coefficient 
in the environment that occur on the scales much less
than those of the observed spatial variations of kinetic processes. 
From Eq.~\eqref{eq:station_sol}
with $f(\varepsilon)=\gamma\varepsilon$ and
$g(\varepsilon)=\varepsilon$,
we obtain the power law
\begin{align}
  \label{eq:power_law}
 \rho_{\st}(\varepsilon) = N/\varepsilon^{\nu}, 
\end{align}
where $\nu=1+\avr{\gamma}/\sigma^2$,
describing the dependence of the stationary distribution 
on the energy.
Note that similar result can be found in Ref.~\cite{Lefever:bk:1984} 
and the above power law distribution significantly 
differs from the well-known solution for the case where the coefficients of
friction and diffusion are independent of the energy. 
% This simple result, can be also obtained in terms of the velocity.

In the case of the ``negative friction'' with $\avr{\gamma}<0$,
the system absorbs the energy and,
for the semi-indefinite energy interval $[\varepsilon_{\min},\infty)$, 
the stationary solution~\eqref{eq:power_law}
does not represent the steady state distribution
 as it fails to meet the normalization condition.
So, in order to have a steady state,
we need to introduce a mechanism 
that limits the energy absorption.
For this purpose, we consider 
the system characterized by the quadratic exchange
function 
$f(\varepsilon)=\avr{\gamma} \varepsilon+\gamma_2 \varepsilon^{2}$,
where the second order term with the positive coefficient
$\gamma_2>0$ bounds the energy absorption from above,
so that $f>0$ at $\varepsilon>\max\{-\avr{\gamma}/\gamma_2,0\}$. 
At $\gamma_2>0$, the power law distribution~\eqref{eq:power_law} 
assumes the following modified form: 
\begin{align}
  \label{eq:power_law_mod}
 \rho_{\st}(\varepsilon) = N  \varepsilon^{-\nu}
\exp\Bigl[
-\frac{\gamma_2}{\sigma^2}\,\varepsilon
\Bigr]. 
\end{align}
Interestingly, 
at $\gamma_2/\sigma^2=\beta$
and $\nu=(2-n)/2$ [$\avr{\gamma}/\sigma^2=-n/2$],
the steady state distribution~\eqref{eq:power_law_mod}
gives the well-known
Maxwell distribution function of Brownian particles
moving in the $n$-dimensional Euclidean space.

In conclusion, let us briefly comment on the so-called 
phenomenological Rayleigh model of active
friction~\cite{Rayleigh:bk:1945}.
In this model, the friction coefficient 
expressed in terms of the velocity 
$\gamma=-\gamma_{0}+\gamma_2 v^{2}\equiv \gamma_2
(v^{2}-v^{2}_{0})=\alpha (\varepsilon-\varepsilon_0)$,
where $v^{2}_{0}=\gamma_{0}/\gamma_2$,
is negative at small velocities, $v^{2}< v^{2}_{0}$,
and, in the low energy region, the system absorbs the energy
from the environment.
By contrast, in the high energy region where
$\varepsilon>\varepsilon_0$,
the friction coefficient is positive and characterizes
the exchange process accompanied by the loss of energy.
In the case of the constant diffusion coefficient
$D_0=\sigma^2 g_0^2$ with $g=g_0$,
the steady state distribution is given by
\begin{align}
  \label{eq:exp_law_Rayleigh}
 \rho_{\st}(\varepsilon) = N 
\exp\Bigl[
-\frac{\alpha}{ 3 D_0}\,
(\varepsilon-\varepsilon_0)^2
(\varepsilon+\varepsilon_0/2)
\Bigr]. 
\end{align}
The only maximum of this distribution
is located at $\varepsilon=\varepsilon_0$
giving the most probable value of the energy
that defines the unique steady state distribution
$\rho_{\st}(\varepsilon)=\delta(\varepsilon-\varepsilon_{0})$
in the limit of low noise $g_0\to 0$. 

%%%%%%%%%%%%%%%%%%%%%%%%%%%%
\section{Entropy and fluctuation relations}
\label{sec:entropy-fluct}
%%%%%%%%%%%%%%%%%%%%%%%%%%%%

%%%%%%%%%%%%%%%%%%%%%%
\subsection{Entropies and effective potential}
\label{subsec:entropy}
%%%%%%%%%%%%%%%%%%%%%

In Sec.~\ref{sec:steady_states} we have found that,
when the stationary probability flux is zero, $J_{\st}=0$,
the steady state distribution~\eqref{eq:station_sol} in the energy representation
is completely determined by the effective energy
potential~\eqref{eq:V_potential},
whereas,
in more general case with non-vanishing flux,
$J_{\st}\ne 0$, the stationary distribution~\eqref{eq:steady_state_gen} 
additionally depends on the diffusion coefficient $D$.

In order to further clarify the role of the effective potential
we consider the trajectory-dependent entropy
of the system~\cite{Crooks:pre:1999,Seifert:prl:2005} 
\begin{align}
  \label{eq:entropy-def}
  s(t)=-\ln \Bigl[
\rho\bigl(\varepsilon,t\bigr)
\Bigr]_{\varepsilon=\varepsilon(t;\xi)},
\end{align}
defined for the trajectory $\varepsilon(t;\xi)$ representing
the noise dependent solution  of the 
Langevin equation~\eqref{eq:Langv_energy}.

Averaging the trajectory-dependent entropy~\eqref{eq:entropy-def}
over noise gives the well-known result
\begin{align}
  \label{eq:entropy-avr}
  \avr{s(t)} = -\int 
\rho(\varepsilon,t)
\ln \bigl[
\rho(\varepsilon,t)
\bigr]\,\dd\varepsilon
\end{align} 
that can be easily obtained using the general relation
\begin{align}
  \label{eq:aux-rel-1}
  \avr{\phi(\varepsilon(t;\xi))}=
\int \avr{\delta(\varepsilon(t;\xi)-\varepsilon)}
\phi(\varepsilon)\,\dd\varepsilon=
\int\rho(\varepsilon,t)\phi(\varepsilon)\,\dd\varepsilon
\equiv
\qsca{\phi}{\rho(t)},
\end{align}
where $\phi$ is a function  of the energy
and integrals are taken over the whole energy range. 
Another useful identity
\begin{align}
  \label{eq:aux-rel-2}
  \avr{\phi(\varepsilon(t;\xi))\dot{\varepsilon}(t;\xi)}=
\int \avr{\delta(\varepsilon(t;\xi)-\varepsilon)\dot{\varepsilon}(t;\xi)}
\phi(\varepsilon)\,\dd\varepsilon=
-\int J(\varepsilon,\rho)\phi(\varepsilon)\,\dd\varepsilon
\equiv
-\bra{\phi}\oprt{J}\ket{\rho}
\end{align}
is at the heart of the derivation of the FP
equation~\eqref{eq:FP_flux}.

In the equation of motion
for the entropy of the system~\eqref{eq:entropy-def}
\begin{align}
  \label{eq:eq_motion}
  \dot{s}(t)=-(\pdrs{t}\rho+\rho^{\,\prime}\dot{\varepsilon})/\rho=
  -\Bigl[
\frac{\pdrs{t}\rho}{\rho} +\frac{J \dot{\varepsilon}}{D \rho} 
\Bigr]
+ V' \dot{\varepsilon}
\end{align}
we may single out 
the contribution due to change in 
entropy of the environment, $s_{\med}$,
related to the rate of heat exchange in the medium
\begin{align}
  \label{eq:entropy_medium}
   \dot{s}_{\med}(t)=-V' \dot{\varepsilon}.
\end{align}
By using the
identity~\eqref{eq:aux-rel-2} it is not difficult
to evaluate its average
\begin{align}
  \label{eq:avr_entropy_medium}
   \avr{\dot{s}_{\med}(t)}=\int J(\varepsilon,\rho)V'(\varepsilon)\,\dd\varepsilon.
\end{align}
Similarly, averaging the increase in
the total entropy $s_{\mathrm{tot}}=s+s_{\med}$
gives the expression for the total entropy
production rate
 \begin{align}
  \label{eq:avr_total_entropy}
  \avr{\dot{s}_{\mathrm{tot}}(t)}=
   \avr{\dot{s}(t)+\dot{s}_{\med}(t)}=\int \frac{J^2(\varepsilon,\rho)}{D(\varepsilon)\rho(\varepsilon,t)}\,\dd\varepsilon
\ge 0
\end{align}
which clearly cannot be negative.
Upon reaching the steady state 
characterized by the stationary distribution~\eqref{eq:steady_state_gen},
the production rates of the total and medium entropies take 
their steady state values given by
 \begin{align}
  \label{eq:total_entropy_st}
&
  \avr{\dot{s}_{\mathrm{tot}}}_{\st}=
   \int
   \frac{J^2_{\st}}{D(\varepsilon)\rho_{\st}(\varepsilon)}\,\dd\varepsilon
 =J_{\st}\ln
\Bigl[
1+J_{\st}/N_{\st}\int\exp\{V(\varepsilon)\}/D(\varepsilon)\,\dd\varepsilon
\Bigr],
\\
\label{eq:medium_entropy_st}
&
  \avr{\dot{s}_{\med}}_{\st}=
J_{\st}\Delta V,
\end{align}
where $\Delta V=V(\varepsilon_{\max})-V(\varepsilon_{\min})$.
As evident from Eqs.~\eqref{eq:total_entropy_st}
and~\eqref{eq:medium_entropy_st}, 
the entropy production rates both 
tend to zero only if the stationary flux vanishes, $J_{\st}=0$.

In the path integral representation, Langevin dynamics
governed by the equation of motion~\eqref{eq:Langv_energy}
is described by the generating functional
of correlation functions, $G[A]$, written
in the functional integral form~\cite{Just:bk4:2002,Vas:bk:2004}.
Applying the standard
procedure~\cite{Hoch:pre:1999,Kis:susy}, 
we deduce the generating functional
\begin{align}
  \label{eq:gnrt_functional}
&
  G[A]=
\int\DD[\varepsilon]
\prod_{t} g^{-1}\exp
\Bigl\{
-S_{\eff}[\varepsilon]+\int_{0}^{t} A(\tau)\varepsilon(\tau)\dd\tau
\Bigr\}
\end{align}
expressed in terms of the effective action
\begin{align}
\label{eq:eff_action}
 &
S_{\eff}[\varepsilon]=\frac{1}{2}
\int_{0}^{t} 
\Bigl\{
\frac{(\dot{\varepsilon}+D V^{\prime})^2}{2D}
-f^{\,\prime}
\Bigr\}
\dd\tau
\end{align}
that determines the weight, $P[\varepsilon]\propto \exp(-S_{\eff})$, of a
trajectory $\varepsilon(\tau)$.
Equation~\eqref{eq:eff_action} agrees with
the results for the effective action previously derived
in Refs~\cite{Arnold:pre:2:2000,Lau:pre:2007,Arenas:pre:2010}.

From the expression~\eqref{eq:eff_action},
it is straightforward to evaluate the difference in the effective action 
for each forward path $\varepsilon(\tau)$
and the corresponding reversed trajectory (backward path), 
$\overline{\varepsilon}(\tau)=\varepsilon(t-\tau)$:
\begin{align}
  \label{eq:delta_S_eff}
  S_{\eff}[\varepsilon]-S_{\eff}[\overline{\varepsilon}]=
\int_{0}^{t}V^{\,\prime}\dot{\varepsilon}\,\dd\tau=-\Delta s_{\med}.
\end{align}
This formula along the medium entropy defined by
the relation~\eqref{eq:entropy_medium}   
shows that this is the entropy generation in the medium,
$\Delta s_{\med}$,
which is solely responsible for
the change in the weight  under ``time reversal'':
$ P[\varepsilon]/P[\overline{\varepsilon}]=\exp(\Delta s_{\med})$.
 
Note that, for suitably defined dissipation function,
this result can be regarded as a version of
the Evans-Searles fluctuation theorem~\cite{Searles:pre:1999}.
It is also applicable to  
externally driven Brownian systems.
In this case, there is a set of external control parameters,
$\lambda(\tau)=\{\lambda_1(\tau),\ldots,\lambda_k(\tau)\}$,
that vary in time 
from $\lambda_0=\lambda(0)$ to $\lambda_t=\lambda(t)$
according to the prescribed (forward) protocol,
whereas the reversed (backward) protocol
is represented by the parameters
$\overline{\lambda}(\tau)=\lambda(t-\tau)$.
The energy potential, $V=V(\varepsilon,\lambda(t))$,  and 
the diffusion coefficient, $D=D(\varepsilon,\lambda(t))$,
are now a function of the time-dependent parameters,
so that the FP and flux operators are both non-stationary.

Despite the evolution operator, $\oprt{U}(t,t_0)$, is no longer given by 
the exponential solution $\oprt{U}(t,t_0)=\exp[-(t-t_0)\oprt{H}]$ 
to the Cauchy problem~\eqref{eq:FP-oper-evol}
with the stationary FP Hamiltonian,
the expression for the effective action~\eqref{eq:eff_action} 
remains intact and its protocol dependent difference, 
$S_{\eff}[\varepsilon,\lambda]-S_{\eff}[\overline{\varepsilon},\overline{\lambda}]$,
is still given by the formula~\eqref{eq:delta_S_eff},
where $V'\equiv V'(\varepsilon,\lambda)$.
This formula and the inequality~\eqref{eq:total_entropy_st}
justify our definition of the entropy of the environment. 

%%%%%%%%%%%%%%%%%%%%%%%%%%%%%
\subsection{Generalized integral fluctuation relation}
\label{subsec:fluc-rels}
%%%%%%%%%%%%%%%%%%%%%%%%%%%%

In general, there are a number of 
relations that can be derived by  making the comparison 
between  the trajectories and reversed ``anti-trajectories''.
Some of these~---~the so-called 
the fluctuation theorems~---~were recently
reviewed in Refs.~\cite{Evans:advph:2002,Marconi:phrep:2008,Esposito:rmp:2009}.
The fluctuation relations were tested experimentally
in a variety of different systems
such as 
colloidal particles manipulated by 
laser
traps~\cite{Wang:prl:2002,Carberry:prl:2004,Trepag:pnas:2004,Wang:pre:2005,
Blickle:prl:2006,Carberry:joa:2007,Andrieux:prl:2007},
biomolecules pulled by AFM's or optical 
tweezers~\cite{Liphardt:sci:2002,Collin:nat:2005},
atomic force microscopy
cantilever~\cite{Solano:epl:2010}
and an electric circuit with an imposed mean current~\cite{Andrieux:prl:2007}.

In this section we 
apply the operator  approach to
deduce
the integral fluctuation relation 
that can be regarded as the generalized
version of 
the well-known Hatano-Sasa identity~\cite{Hatano:prl:2001}.
We also discuss how to recover other known results
such as the Jarzynski's equality~\cite{Jarzyn:prl:1997}
and the relation for the total entropy
obtained by Seifert in Ref.~\cite{Seifert:prl:2005}.
Note that the fluctuation relations for 
stochastic systems originally studied in 
Refs.~\cite{Kurchan:jpa:1998,Lebowitz:jsp:1999,Maes:jsp:1999,Searles:pre:1999}
were recently extended to the case of inhomogeneous stochastic
processes~\cite{Ge:jpa:2007,Ge:jsp:2008,Shargel:jpa:2010}.

Our first step is to introduce
a family of
the modified FP operators 
\begin{align}
\label{eq:Hp}
&
   \oprt{H}_p=
-\pdrs{\varepsilon}\cdot\oprt{J}_p,
\quad
\oprt{J}_p= \oprt{J} - p/\rho_{\st},  
\end{align}
where $p$ is a real number which might be called  
the \textit{flux parameter}.
From Eq.~\eqref{eq:Hp}  it can be seen that  
the operator~\eqref{eq:FP-Hamilt-flux},
$\oprt{H}=\oprt{H}_0$,
transforms into $\oprt{H}_p$
with the shifted flux operator $\oprt{J}_p$
when the energy derivative of the potential
$V'$ is replaced with $V'-p/(D\rho_{\st})$. 
So, the energy potential, $V_p$, for the deformed FP operator
$\oprt{H}_p$ is given by
\begin{align}
  \label{eq:Vp}
  V_p =V -\frac{p}{J_{\st}}\,\ln
\Bigl[
1+J_{\st}/N_{\st}\int_{\varepsilon_0}^{\varepsilon}\exp\{V(\varepsilon')\}/D(\varepsilon')\,\dd\varepsilon'
\Bigr],
\end{align}
where the potential $V$ is defined in Eq.~\eqref{eq:V_potential}.

It is straightforward to check
the validity of the algebraic identity
\begin{align}
   \label{eq:Herm_conj-rel}
&
\hcnj{\oprt{H}}_p
=\hcnj{\oprt{J}}_{p}\cdot\pdrs{\varepsilon}=
\ee^{\Psi_{\st}}\cdot\oprt{H}_q\cdot\ee^{-\Psi_{\st}},
\quad
q=2 J_{\st}-p,
\\
\label{eq:Psi_st}
&
\Psi_{\st}=-\ln \rho_{\st},
 \end{align}
where
the superscript $\dagger$ stands for Hermitian conjugation,
linking the Hermitian conjugate of the deformed FP operator~\eqref{eq:Hp},
$\hcnj{\oprt{H}}_p$, at $p=J_{\st}+\Delta J$
and the operator $\oprt{H}_q$ with the flux parameter
$q=2 J_{\st}-p=J_{\st}-\Delta J$ through the steady state potential
 $\Psi_{\st}$, of the stationary distribution, $\rho_{\st}$,
characterized by the flux number (stationary probability flux)
$J_{\st}$.  

Another important point
is that
the evolution operator $\oprt{U}_{p}(t,0;\overline{\lambda})$ 
of the deformed Hamiltonian~\eqref{eq:Hp}
computed at the reversed protocol $\overline{\lambda}$
preserves the normalization
condition of a probability distribution function
$\rho_{\mathrm{f}}$: 
$\qsca{1}{\rho_{\mathrm{f}}}=
\qsca{\rho_{\mathrm{f}}}{1}=\int\rho_{\mathrm{f}}(\varepsilon)\,\dd\varepsilon=1$.
Our method is to combine 
the normalization preserving condition
\begin{align}
  \label{eq:norm_aux}
  \bra{1}\oprt{U}_{p}(t,0;\overline{\lambda})\ket{\rho_{\mathrm{f}}}
= \bra{\rho_{\mathrm{f}}}\hcnj{\oprt{U}}_{p}(t,0;\overline{\lambda})\ket{1}
=\qsca{\rho_{\mathrm{f}}}{1}
=1
\end{align}
with the identity~\eqref{eq:Herm_conj-rel}.
To this end, we slice 
the time interval $[0,t]$ into a large number $N+1$
of small pieces of the thickness $\Delta\tau=t/(N+1)$
and approximate the evolution operator
$\oprt{U}_{p}(\tau_i,\tau_{i-1};\overline{\lambda})$,
where $\tau_i=\tau_{i-1}+\Delta\tau$ and $\tau_0=0$,
by the operator exponent  
$\exp[-\Delta\tau \oprt{H}_p(\overline{\lambda}_i)]$,
where 
$
\overline{\lambda}_i\equiv
\overline{\lambda}(\tau_i)=
\lambda_{N+1-i}$.
Then, by using the identity~\eqref{eq:Herm_conj-rel}, 
the discretized evolution operator
\begin{align}
  \label{eq:U_discr}
  \hcnj{\oprt{U}}_{p}(t,0;\overline{\lambda})
\approx
\hcnj{
\Bigl[
\ee^{-\Delta\tau\oprt{H}_p(\lambda_0)}\cdot\ee^{-\Delta\tau\oprt{H}_p(\lambda_1)}\cdots
\ee^{-\Delta\tau\oprt{H}_p(\lambda_N)}
\Bigr]
}.
\end{align}
can be recast into the following operator product 
\begin{align}
  \label{eq:Up_Uq}
  \hcnj{\oprt{U}}_{p}(t,0;\overline{\lambda})
\approx
\rho_{N}^{-1}\cdot
\ee^{-\Delta\tau\oprt{H}_q(\lambda_N)}\cdot\ee^{-\Delta\Psi_N}\cdots
\ee^{-\Delta\tau\oprt{H}_q(\lambda_1)}\cdot
\ee^{-\Delta\Psi_1}\cdot
\ee^{-\Delta\tau\oprt{H}_q(\lambda_0)}
\cdot\rho_{0},
\end{align}
where $\rho_i=\rho_{\st}(\lambda_i)$ and $\exp[-\Delta\Psi_i]=\rho_i/\rho_{i-1}$.

Assuming that 
the initial probability distribution
is $\rho_{\mathrm{in}}$,
we derive the equality
\begin{align}
  \label{eq:gen-HS-discr}
&
\bra{1}
\ee^{-(\Psi_{\mathrm{f}}-\Psi_N)}\cdot
\ee^{-\Delta\tau\oprt{H}_q(\lambda_N)}\cdot\ee^{-\Delta\Psi_N}\cdots
\ee^{-\Delta\tau\oprt{H}_q(\lambda_1)}\cdot
\ee^{-\Delta\Psi_1}\cdot
\ee^{-\Delta\tau\oprt{H}_q(\lambda_0)}\cdot
\ee^{\Psi_{\mathrm{in}}-\Psi_0}
\ket{\rho_{\mathrm{in}}}=1,  
\end{align}
where
$\Psi_{\mathrm{in,\, f}}=-\ln\rho_{\mathrm{in,\, f}}$,
which, in the limit of large slice number,
$N\to\infty$,
gives our key result in the form of the identity
\begin{align}
  \label{eq:gen-HS-idnt}
&
%\xrightarrow{N\to\infty}
\biggl\langle
\exp
\Bigl\{
-\int_{0}^{t}
\pdr{\Psi_{\st}}{\lambda_\alpha}\,\dot{\lambda}_\alpha\,
\dd\tau
+\Delta\Psi_{\mathrm{in}}-\Delta\Psi_{\mathrm{f}}
\Bigr\}
\biggr\rangle_{q}
=1,
\end{align}
where
$\Delta\Psi_{\mathrm{in}}=\Psi_{\mathrm{in}}-\Psi_{\st}(\lambda_0)=-\ln[\rho_{\mathrm{in}}/\rho_{\st}(\lambda_0)]$
and
$\Delta\Psi_{\mathrm{f}}=\Psi_{\mathrm{f}}-\Psi_{\st}(\lambda_t)=-\ln[\rho_{\mathrm{f}}/\rho_{\st}(\lambda_t)]$.
and the index $q$ indicates using  the deformed FP operator $\oprt{H}_q$. 

The relation~\eqref{eq:gen-HS-idnt} involves 
the flux parameter, $q$, and
the two probability distributions,  $\rho_{\mathrm{f}}$
and $\rho_{\mathrm{in}}$. 
In what follows we concentrate on 
the important case of non-deformed effective potential, 
where the flux parameter $q$ is zero and $p=2J_{\st}$.
On substituting
$\rho_{\mathrm{in}}=\rho_{\st}(\lambda_0)$
and $\rho_{\mathrm{f}}=\rho_{\st}(\lambda_t)$
into the identity~\eqref{eq:gen-HS-idnt},
we recover the result
obtained by 
Hatano and Sasa in Ref.~\cite{Hatano:prl:2001} 
\begin{align}
  \label{eq:HS-ident}
  \biggl\langle
\exp
\Bigl\{
-\int_{0}^{t}
\pdr{\Psi_{\st}}{\lambda_\alpha}\,\dot{\lambda}_\alpha\,
\dd\tau
\Bigr\}
\biggr\rangle
=1
\end{align}
which is just the  special case of the relation~\eqref{eq:gen-HS-idnt}
with $q=0$ and $\Delta\Psi_{\mathrm{in},\,\mathrm{f}}=0$.

When the stationary flux vanishes and $J_{\st}=0$,
the steady state distribution is given in
Eq.~\eqref{eq:station_sol} with 
$\Psi_{\st}=V-F_{\st}$,
so that the Hatano-Sasa formula~\eqref{eq:HS-ident}
can be rewritten as
the Jarzynski's equality~\cite{Jarzyn:prl:1997}
\begin{align}
  \label{eq:Jarz}
  \biggl\langle
\exp
\Bigl\{
-\int_{0}^{t}
\pdr{V}{\lambda_\alpha}\,\dot{\lambda}_\alpha\,
\dd\tau
\Bigr\}
\biggr\rangle
=\exp\{-\Delta F_{\st}\},
\end{align}
where $\Delta F_{\st}=F_{\st}(\lambda_t)- F_{\st}(\lambda_0)$.

Since the integral
\begin{align}
  \label{eq:Delta_V}
  \int_{0}^{t}
\pdr{V}{\lambda_\alpha}\,\dot{\lambda}_\alpha\,
\dd\tau=\Delta V-
  \int_{0}^{t}
V' \,\dot{\varepsilon}\,
\dd\tau=\Delta V +\Delta s_{\med}
\end{align}
can be expressed
in terms of the entropy of the medium~\eqref{eq:entropy_medium},
at $J_{\st}=0$,
the relation~\eqref{eq:gen-HS-idnt} takes the following form
\begin{align}
  \label{eq:Delta_s_m}
  \Bigl\langle
\exp
\bigl\{
-\Delta s_{\med}-(\Psi_{\mathrm{f}}-\Psi_{\mathrm{in}})
\bigr\}
\Bigr\rangle
=1
\end{align}
that reduces to the integral fluctuation theorem
for the total change in entropy~\cite{Seifert:prl:2005,Speck:jsm:2007,Seifert:epjb:2008}
\begin{align}
  \label{eq:Delta_s_tot}
  \Bigl\langle
\exp
\bigl\{
-\Delta s_{\mathrm{tot}}
\bigr\}
\Bigr\rangle
=1
\end{align}
when
$\ket{\rho_{\mathrm{f}}}=\oprt{U}(t,0)\ket{\rho_{\mathrm{in}}}$
and $\Delta s = \Psi_{\mathrm{f}}-\Psi_{\mathrm{in}}$.
From Eq.~\eqref{eq:Delta_s_tot} combined with the Jensen's inequality 
$\avr{\exp(x)}\ge \exp\avr{x}$,
the averaged change in the total
entropy cannot be negative, so that, 
in agreement with Eq.~\eqref{eq:avr_total_entropy}  and the second law
of thermodynamics,
$\avr{\Delta s_{\mathrm{tot}}}\ge 0$.

So, we have found that 
the well-known results such as
the Hatano-Sasa identity~\eqref{eq:HS-ident},
the Jarzynski's equality~\eqref{eq:Jarz}
and the fluctuation theorem for the total entropy~\eqref{eq:Delta_s_tot}
immediately follow from
our generalized fluctuation relation~\eqref{eq:gen-HS-idnt}
derived for nonzero flux parameters. 

Our concluding remark 
concerns
the generalized fluctuation-exchange
theorem for the steady-state systems
formulated in a very recent paper~\cite{Prost:prl:2009}.
This theorem is essentially a direct consequence
of the Hatano-Sasa identity~\eqref{eq:HS-ident}
applied to  the limiting case of small perturbations,
where $\lambda(t)=\lambda_0+\delta \lambda(t)$.
More specifically,  it asserts that
the response functions
\begin{align}
   \label{eq:respn-func}
&
\frac{
\delta
  \avr{\Psi_{\alpha}(t)}
}{
\delta\lambda_{\beta}(\tau)
}
=
R_{\alpha\beta}(t-\tau)
\end{align}
of the averages
\begin{align}
\label{eq:Psi_vs_t}
&
\avr{\Psi_{\alpha}(t)}=
\bra{1}\Psi_{\alpha}\cdot\oprt{U}(t,0)\ket{\rho_0},
\quad
% \\
% &
%   \label{eq:Psi_alp}
  \Psi_{\alpha}\equiv\pdr{\Psi_{\st}}{\lambda_\alpha}\biggr|_{\lambda=\lambda_0},
\quad
\rho_0\equiv\rho_{\st}\bigr|_{\lambda=\lambda_0}
\end{align}
meet the fluctuation-exchange relation
\begin{align}
  \label{eq:FDT}
&
R_{\alpha\beta}(t-\tau)=
\pdrs{t}C_{\alpha\beta}(t-\tau),
\end{align}
where $C_{\alpha\beta}(t-\tau)$ is the correlation function given by
\begin{align}
\label{eq:C_ab}
& 
C_{\alpha\beta}(t-\tau)=
\avr{\Psi_{\alpha}(t)\Psi_{\beta}(\tau)}_0=
\bra{1}\Psi_{\alpha}\cdot
\exp[-(t-\tau)\oprt{H}(\lambda_0)]
\cdot
\Psi_{\beta}\ket{\rho_0}. 
\end{align}
From the above considerations, the 
formulas~\eqref{eq:respn-func}--~\eqref{eq:C_ab} 
written down explicitly in the bracket notations
are applicable to the Brownian systems 
in the energy representation.
We shall extend on the subject elsewhere.

%%%%%%%%%%%%%%%%%%%%%%%%%%%%
\section{Discussion and conclusions}
\label{sec:discuss}
%%%%%%%%%%%%%%%%%%%%%%%%%%%

In this paper we have demonstrated that 
the  method that combines  
the energy representation
with the Langevin dynamics approach 
may provide an effective tool to explore
the steady states of certain nonequilibrium 
Brownian systems.
These states can be analytically studied as the stationary
solutions of the corresponding FP equation
expressed in terms of 
the exchange and diffusion functions,
$f(\varepsilon)$ and $g(\varepsilon)$.

The function of energy exchange rate, $f$, 
defined as the deterministic part
of the Langevin equation~\eqref{eq:Langv_energy} 
describes the process of direct energy  interchange between the system
and the environment,
whereas 
the multiplicative white noise
characterized by the diffusion function $g$
represent the fluctuation induced effects.
The latter may depend on the energy of the system
and hence the diffusion function is generally not a constant. 

We have shown that,
in the limit of low noise with $g\to 0$,
the steady states are determined by equilibria of 
the effective potential, $\tilde{E}$, 
defined in Eq.~\eqref{eq:Cauchy_prb}.
At $g=g_0\ne 0$, these equilibria correspond to the maxima of 
the steady state (stationary) distribution, $\rho_{\st}$~\eqref{eq:station_sol}. 

In the well-known case of
linear exchange function, $f=\gamma(\varepsilon-\varepsilon_0)$,
the system losses its energy via 
the processes of dissipation  when the friction coefficient is positive
$\gamma>0$.
Then the energy $\varepsilon_0$ is 
the only equilibrium point of the harmonic potential, 
$\tilde{E}=\gamma(\varepsilon-\varepsilon_0)^2/2$, and, 
at energy independent diffusion function $g=g_0$,
the  steady state distribution is of the Gaussian form:
$\rho_{\st}= N \exp\bigl[-\gamma
(\varepsilon-\varepsilon_0)^2/(2 D_0)\bigr]$.

In the opposite case of negative friction coefficient
with $\gamma<0$,
the system absorbs the energy from the environment 
and there are no  steady states without 
mechanisms limiting the gain of energy. 
Interestingly, we have found that,
when the friction coefficient
depends on the energy and contains 
the random contribution 
coming from the environment stochasticity, 
the steady state distribution
may take the form of the standard Maxwell equilibrium
distribution for Brownian particles
(the conditions can be found  below Eq.~\eqref{eq:power_law_mod}).

So, using our approach it is relatively easy to conclude
on stationary distribution functions of nonequilibrium Brownian 
systems for different mechanisms of  energy interchange 
complicated by the stochastic nonlinear coupling 
with the environment.
In particular, 
it is not difficult to recover
the results on statistics of energy states
obtained in Ref.~\cite{Tuzel:jsp:2000}
for a generic model of  a random polypeptide chain
that reproduces the energy probability distribution
of real proteins over a very large range of energies. 
It  is also pertinent to note that
the systems such as dust particles in plasma~\cite{Lev:pla:2009} 
and Brownian particles 
with the randomly inhomogeneous friction coefficient 
present important cases for which
theoretical predictions can be experimentally verified,
but this task requires additional analysis and 
experimental data.

As first steps in this direction,
we have introduced the trajectory-dependent entropy
in the energy representation
so as to define both the entropy of the Brownian system 
and the entropy production in 
the environment. 
The latter, similar to the steady state distributions at $J_{\st}=0$, 
was found to be determined
by the effective energy potential~\eqref{eq:V_potential}.

As far as the fluctuation relations are concerned,
our key finding is
the generalized integral fluctuation theorem~\eqref{eq:gen-HS-idnt}
describing nonequilibrium Brownian systems
with the shifted effective potential
characterized by the flux parameter, $q$, and
the flux number, $J_{\st}$ 
(the stationary value of the probability current).
It turned out  that the fluctuation theorem for the total entropy change~\eqref{eq:Delta_s_tot}
can be deduced from this theorem.
The Hatano-Sasa identity~\eqref{eq:HS-ident}
and
the Jarzynski's equality~\eqref{eq:Jarz}
can equally be derived as the special cases of our fluctuation
relation~\eqref{eq:gen-HS-idnt}.

At $p=J_{\st}=0$, 
the algebraic identity~\eqref{eq:Herm_conj-rel} reduces to the detailed
balance relation
\begin{align}
  \label{eq:detail_balance}
 \hcnj{\oprt{H}}=
\ee^{\Psi_{\st}}\cdot\oprt{H}\cdot\ee^{-\Psi_{\st}}  
\end{align}
for the steady state distribution of the potential form~\eqref{eq:station_sol}.
In the opposite case characterized by nonzero flux number, $J_{\st}\ne 0$,
and non-reflecting boundary conditions, 
the identity~\eqref{eq:Herm_conj-rel} describes 
the systems where detailed balance is violated.
Alternatively, in multidimensional systems, explicit violation of detailed balance
can be caused by non-conservative forces~\cite{Chernyak:jsm:2006}.

From the integral theorem~\eqref{eq:gen-HS-idnt},
and, following the line of reasoning presented in Ref.~\cite{Marin:epl:2008},
it can be concluded that
the energy representation serves as a coarse-grained description
of stochastic systems where the energy can be viewed as the
reduced variable such that information on its trajectories
is enough to reproduce the statistics of the entropy production.

Our concluding remark concerns the detailed theorem
that underlies the integral fluctuation relation~\eqref{eq:gen-HS-idnt}. 
According to the general results of the very recent
paper~\cite{Esposito:prl:2010},
 when a variable obeys an integral fluctuation theorem,
it automatically obeys a detailed theorem.
So, the relation~\eqref{eq:gen-HS-idnt}. additionally idicates that,
in the energy representation, 
there are detailed fluctuation theorems describing
Brownian-like systems in the absence of  detailed balance.
These theorems and related analysis are beyond the scope of this
paper. 
We shall extend on this subject elsewhere.

%\bibliography{optics,disord,polymer,scatter,lc,quant,hk,flc,qft,math}

\begin{thebibliography}{63}%
\makeatletter
\providecommand \@ifxundefined [1]{%
 \@ifx{#1\undefined}
}%
\providecommand \@ifnum [1]{%
 \ifnum #1\expandafter \@firstoftwo
 \else \expandafter \@secondoftwo
 \fi
}%
\providecommand \@ifx [1]{%
 \ifx #1\expandafter \@firstoftwo
 \else \expandafter \@secondoftwo
 \fi
}%
\providecommand \natexlab [1]{#1}%
\providecommand \enquote  [1]{``#1''}%
\providecommand \bibnamefont  [1]{#1}%
\providecommand \bibfnamefont [1]{#1}%
\providecommand \citenamefont [1]{#1}%
\providecommand \href@noop [0]{\@secondoftwo}%
\providecommand \href [0]{\begingroup \@sanitize@url \@href}%
\providecommand \@href[1]{\@@startlink{#1}\@@href}%
\providecommand \@@href[1]{\endgroup#1\@@endlink}%
\providecommand \@sanitize@url [0]{\catcode `\\12\catcode `\$12\catcode
  `\&12\catcode `\#12\catcode `\^12\catcode `\_12\catcode `\%12\relax}%
\providecommand \@@startlink[1]{}%
\providecommand \@@endlink[0]{}%
\providecommand \url  [0]{\begingroup\@sanitize@url \@url }%
\providecommand \@url [1]{\endgroup\@href {#1}{\urlprefix }}%
\providecommand \urlprefix  [0]{URL }%
\providecommand \Eprint [0]{\href }%
\@ifxundefined \urlstyle {%
  \providecommand \doi  [0]{\begingroup \@sanitize@url \@doi}%
  \providecommand \@doi [1]{\endgroup \@@startlink {\doibase
  #1}doi:\discretionary {}{}{}#1\@@endlink }%
}{%
  \providecommand \doi  [0]{doi:\discretionary{}{}{}\begingroup
  \urlstyle{rm}\Url }%
}%
\providecommand \doibase [0]{http://dx.doi.org/}%
\providecommand \Doi [0]{\begingroup \@sanitize@url \@Doi }%
\providecommand \@Doi  [1]{\endgroup\@@startlink{\doibase#1}\@@Doi}%
\providecommand \@@Doi [1]{#1\@@endlink}%
\providecommand \selectlanguage [0]{\@gobble}%
\providecommand \bibinfo  [0]{\@secondoftwo}%
\providecommand \bibfield  [0]{\@secondoftwo}%
\providecommand \translation [1]{[#1]}%
\providecommand \BibitemOpen [0]{}%
\providecommand \bibitemStop [0]{}%
\providecommand \bibitemNoStop [0]{.\EOS\space}%
\providecommand \EOS [0]{\spacefactor3000\relax}%
\providecommand \BibitemShut  [1]{\csname bibitem#1\endcsname}%
%</preamble>
\bibitem [{\citenamefont {Huang}(1987)}]{Huang:bk:1987}%
  \BibitemOpen
  \bibfield  {author} {\bibinfo {author} {\bibfnamefont {K.}~\bibnamefont
  {Huang}},\ }\href@noop {} {{\selectlanguage {english}\emph {\bibinfo {title}
  {Statistical Mechanics}}}},\ \bibinfo {edition} {2nd}\ ed.\ (\bibinfo
  {publisher} {Wiley},\ \bibinfo {address} {NY},\ \bibinfo {year} {1987})\ p.\
  \bibinfo {pages} {506}\BibitemShut {NoStop}%
\bibitem [{\citenamefont {Huang}(2005)}]{Huang:bk:2005}%
  \BibitemOpen
  \bibfield  {author} {\bibinfo {author} {\bibfnamefont {K.}~\bibnamefont
  {Huang}},\ }\href@noop {} {{\selectlanguage {english}\emph {\bibinfo {title}
  {Lectures on Statistical Physics and Protein Folding}}}}\ (\bibinfo
  {publisher} {World Scientific},\ \bibinfo {address} {Singapore},\ \bibinfo
  {year} {2005})\ p.\ \bibinfo {pages} {144}\BibitemShut {NoStop}%
\bibitem [{\citenamefont {Balescu}(1975)}]{Balescu:bk:1975}%
  \BibitemOpen
  \bibfield  {author} {\bibinfo {author} {\bibfnamefont {R.}~\bibnamefont
  {Balescu}},\ }\href@noop {} {{\selectlanguage {english}\emph {\bibinfo
  {title} {Equilibrium and Nonequilibrium Statistical Mechanics}}}}\ (\bibinfo
  {publisher} {Wiley},\ \bibinfo {address} {NY},\ \bibinfo {year} {1975})\ p.\
  \bibinfo {pages} {742}\BibitemShut {NoStop}%
\bibitem [{\citenamefont {Groot}\ and\ \citenamefont
  {Mazur}(1974)}]{Groot:bk:1974}%
  \BibitemOpen
  \bibfield  {author} {\bibinfo {author} {\bibfnamefont {S.~R.~De}\
  \bibnamefont {Groot}}\ and\ \bibinfo {author} {\bibfnamefont
  {P.}~\bibnamefont {Mazur}},\ }\href@noop {} {{\selectlanguage {english}\emph
  {\bibinfo {title} {Non-equilibrium Thermodynamics}}}}\ (\bibinfo  {publisher}
  {Dover},\ \bibinfo {address} {NY},\ \bibinfo {year} {1974})\ p.\ \bibinfo
  {pages} {514}\BibitemShut {NoStop}%
\bibitem [{\citenamefont {Balescu}(1997)}]{Balescu:bk:1997}%
  \BibitemOpen
  \bibfield  {author} {\bibinfo {author} {\bibfnamefont {R.}~\bibnamefont
  {Balescu}},\ }\href@noop {} {{\selectlanguage {english}\emph {\bibinfo
  {title} {Statistical Dynamics: Matter out of Equilibrium}}}}\ (\bibinfo
  {publisher} {Imperial College Press},\ \bibinfo {address} {London},\ \bibinfo
  {year} {1997})\ p.\ \bibinfo {pages} {323}\BibitemShut {NoStop}%
\bibitem [{\citenamefont {Mazenko}(2006)}]{Mazenko:bk:2006}%
  \BibitemOpen
  \bibfield  {author} {\bibinfo {author} {\bibfnamefont {G.~F.}\ \bibnamefont
  {Mazenko}},\ }\href@noop {} {{\selectlanguage {english}\emph {\bibinfo
  {title} {Nonequilibrium Statistical Mechanics}}}}\ (\bibinfo  {publisher}
  {Wiley-VCH},\ \bibinfo {address} {NY},\ \bibinfo {year} {2006})\ p.\ \bibinfo
  {pages} {478}\BibitemShut {NoStop}%
\bibitem [{\citenamefont {Zwanzig}(2001)}]{Zwanzig:bk:2001}%
  \BibitemOpen
  \bibfield  {author} {\bibinfo {author} {\bibfnamefont {R.}~\bibnamefont
  {Zwanzig}},\ }\href@noop {} {{\selectlanguage {english}\emph {\bibinfo
  {title} {Nonequilibrium Statistical Mechanics}}}}\ (\bibinfo  {publisher}
  {Oxford Univ. Press},\ \bibinfo {address} {NY},\ \bibinfo {year} {2001})\ p.\
  \bibinfo {pages} {222}\BibitemShut {NoStop}%
\bibitem [{\citenamefont {Streater}(1995)}]{Streater:bk:1995}%
  \BibitemOpen
  \bibfield  {author} {\bibinfo {author} {\bibfnamefont {R.~F.}\ \bibnamefont
  {Streater}},\ }\href@noop {} {{\selectlanguage {english}\emph {\bibinfo
  {title} {Statistical Dynamics: A Stochastic Approach to Nonequilibrium
  Thermodynamics}}}}\ (\bibinfo  {publisher} {Imperial College Press},\
  \bibinfo {address} {London},\ \bibinfo {year} {1995})\ p.\ \bibinfo {pages}
  {275}\BibitemShut {NoStop}%
\bibitem [{\citenamefont {Ross}(2008)}]{Ross:bk:2008}%
  \BibitemOpen
  \bibfield  {author} {\bibinfo {author} {\bibfnamefont {J.}~\bibnamefont
  {Ross}},\ }\href@noop {} {{\selectlanguage {english}\emph {\bibinfo {title}
  {Thermodynamics and Fluctuations far from Equilibrium}}}},\ Springer Series
  in Chemical Physics\ (\bibinfo  {publisher} {Springer},\ \bibinfo {address}
  {Berlin},\ \bibinfo {year} {2008})\ p.\ \bibinfo {pages} {209}\BibitemShut
  {NoStop}%
\bibitem [{\citenamefont {Nelson}(2001)}]{Nelson:bk:2001}%
  \BibitemOpen
  \bibfield  {author} {\bibinfo {author} {\bibfnamefont {E.}~\bibnamefont
  {Nelson}},\ }\href@noop {} {{\selectlanguage {english}\emph {\bibinfo {title}
  {Dynamical Theories of Brownian Motion}}}},\ \bibinfo {edition} {2nd}\ ed.\
  (\bibinfo  {publisher} {Princeton Univ. Press},\ \bibinfo {address}
  {Princeton},\ \bibinfo {year} {2001})\ p.\ \bibinfo {pages} {114}\BibitemShut
  {NoStop}%
\bibitem [{\citenamefont {Mazo}(2002)}]{Mazo:bk:2002}%
  \BibitemOpen
  \bibfield  {author} {\bibinfo {author} {\bibfnamefont {R.~M.}\ \bibnamefont
  {Mazo}},\ }\href@noop {} {{\selectlanguage {english}\emph {\bibinfo {title}
  {Brownian Motion: Fluctuations, Dynamics, and Applications}}}},\ The
  international series of monographs on physics\ (\bibinfo  {publisher}
  {Claredon Press},\ \bibinfo {address} {Oxford},\ \bibinfo {year} {2002})\ p.\
  \bibinfo {pages} {290}\BibitemShut {NoStop}%
\bibitem [{\citenamefont {Derrida}(1980)}]{Derrida:prl:1980}%
  \BibitemOpen
  \bibfield  {author} {\bibinfo {author} {\bibfnamefont {B.}~\bibnamefont
  {Derrida}},\ }\bibfield  {title} {{\selectlanguage {english}\enquote
  {\bibinfo {title} {Random-energy model: {L}imit of a family of disordered
  models},}\ }}\href@noop {} {\bibfield  {journal} {\bibinfo  {journal} {Phys.
  Rev. Lett.},\ }\textbf {\bibinfo {volume} {45}},\ \bibinfo {pages} {79--82}
  (\bibinfo {year} {1980})}\BibitemShut {NoStop}%
\bibitem [{\citenamefont {Derrida}(1981)}]{Derrida:prb:1981}%
  \BibitemOpen
  \bibfield  {author} {\bibinfo {author} {\bibfnamefont {B.}~\bibnamefont
  {Derrida}},\ }\bibfield  {title} {{\selectlanguage {english}\enquote
  {\bibinfo {title} {Random-energy model: {A}n exactly solvable model of
  disordered systems},}\ }}\href@noop {} {\bibfield  {journal} {\bibinfo
  {journal} {Phys. Rev. B},\ }\textbf {\bibinfo {volume} {24}},\ \bibinfo
  {pages} {2613--2626} (\bibinfo {year} {1981})}\BibitemShut {NoStop}%
\bibitem [{\citenamefont {Derrida}(1985)}]{Derrida:jphq:1985}%
  \BibitemOpen
  \bibfield  {author} {\bibinfo {author} {\bibfnamefont {B.}~\bibnamefont
  {Derrida}},\ }\bibfield  {title} {{\selectlanguage {english}\enquote
  {\bibinfo {title} {A generalization of the random enegy model which includes
  correlations between energies},}\ }}\href@noop {} {\bibfield  {journal}
  {\bibinfo  {journal} {J. Physique Lett.},\ }\textbf {\bibinfo {volume}
  {46}},\ \bibinfo {pages} {L401--L407} (\bibinfo {year} {1985})}\BibitemShut
  {NoStop}%
\bibitem [{\citenamefont {Dyre}(1995)}]{Dyre:prb:1995}%
  \BibitemOpen
  \bibfield  {author} {\bibinfo {author} {\bibfnamefont {J.~C.}\ \bibnamefont
  {Dyre}},\ }\bibfield  {title} {{\selectlanguage {english}\enquote {\bibinfo
  {title} {Energy master equation: {A} low-temperature approximation for
  {B}\"{a}ssler's random-walk model},}\ }}\href@noop {} {\bibfield  {journal}
  {\bibinfo  {journal} {Phys. Rev. B},\ }\textbf {\bibinfo {volume} {51}},\
  \bibinfo {pages} {12276--12294} (\bibinfo {year} {1995})}\BibitemShut
  {NoStop}%
\bibitem [{\citenamefont {T\"uzel}\ and\ \citenamefont
  {Erzan}(2000)}]{Tuzel:jsp:2000}%
  \BibitemOpen
  \bibfield  {author} {\bibinfo {author} {\bibfnamefont {E.}~\bibnamefont
  {T\"uzel}}\ and\ \bibinfo {author} {\bibfnamefont {A.}~\bibnamefont
  {Erzan}},\ }\bibfield  {title} {{\selectlanguage {english}\enquote {\bibinfo
  {title} {Dissipative dynamics and the statistics of energy states of a
  {H}ookean model for protein folding},}\ }}\href@noop {} {\bibfield  {journal}
  {\bibinfo  {journal} {J. Stat. Phys.},\ }\textbf {\bibinfo {volume} {100}},\
  \bibinfo {pages} {405--422} (\bibinfo {year} {2000})}\BibitemShut {NoStop}%
\bibitem [{\citenamefont {Snook}(2007)}]{Snook:bk:2007}%
  \BibitemOpen
  \bibfield  {author} {\bibinfo {author} {\bibfnamefont {I.}~\bibnamefont
  {Snook}},\ }\href@noop {} {{\selectlanguage {english}\emph {\bibinfo {title}
  {The Langevin and Generalised Langevin Approach to The Dynamics of Atomic,
  Polymeric and Colloidal Systems}}}}\ (\bibinfo  {publisher} {Elsevier},\
  \bibinfo {address} {Amsterdam},\ \bibinfo {year} {2007})\ p.\ \bibinfo
  {pages} {303}\BibitemShut {NoStop}%
\bibitem [{\citenamefont {Coffey}\ \emph {et~al.}(2004)\citenamefont {Coffey},
  \citenamefont {Kalmykov},\ and\ \citenamefont {Waldron}}]{Coffey:bk:2004}%
  \BibitemOpen
  \bibfield  {author} {\bibinfo {author} {\bibfnamefont {W.~T.}\ \bibnamefont
  {Coffey}}, \bibinfo {author} {\bibfnamefont {Yu.~P.}\ \bibnamefont
  {Kalmykov}}, \ and\ \bibinfo {author} {\bibfnamefont {J.~T.}\ \bibnamefont
  {Waldron}},\ }\href@noop {} {{\selectlanguage {english}\emph {\bibinfo
  {title} {The Langevin Equation: With Applications to Stochastic Problems in
  Physics, Chemistry and Electrical Engineering}}}},\ \bibinfo {edition} {2nd}\
  ed.,\ \bibinfo {series} {World Scientific Series in Contemporary Chemical
  Physics}, Vol.~\bibinfo {volume} {14}\ (\bibinfo  {publisher} {World
  Scientific},\ \bibinfo {address} {Singapore},\ \bibinfo {year} {2004})\ p.\
  \bibinfo {pages} {678}\BibitemShut {NoStop}%
\bibitem [{\citenamefont {Zinn-Justin}(2001)}]{Just:bk4:2002}%
  \BibitemOpen
  \bibfield  {author} {\bibinfo {author} {\bibfnamefont {J.}~\bibnamefont
  {Zinn-Justin}},\ }\href@noop {} {{\selectlanguage {english}\emph {\bibinfo
  {title} {Quantum Field Theory and Critical Phenomena}}}},\ \bibinfo {edition}
  {4th}\ ed.\ (\bibinfo  {publisher} {Claredon Press},\ \bibinfo {address}
  {Oxford},\ \bibinfo {year} {2001})\ p.\ \bibinfo {pages} {1054}\BibitemShut
  {NoStop}%
\bibitem [{\citenamefont {Gardiner}(2004)}]{Gard}%
  \BibitemOpen
  \bibfield  {author} {\bibinfo {author} {\bibfnamefont {C.~W.}\ \bibnamefont
  {Gardiner}},\ }\href@noop {} {{\selectlanguage {english}\emph {\bibinfo
  {title} {Handbook of Stochastic Methods for Physics, Chemistry and the
  Natural Sciences}}}},\ \bibinfo {edition} {3rd}\ ed.,\ Springer Series in
  Synergetics\ (\bibinfo  {publisher} {Springer -- Verlag},\ \bibinfo {address}
  {Berlin},\ \bibinfo {year} {2004})\BibitemShut {NoStop}%
\bibitem [{\citenamefont {Lau}\ and\ \citenamefont
  {Lubensky}(2007)}]{Lau:pre:2007}%
  \BibitemOpen
  \bibfield  {author} {\bibinfo {author} {\bibfnamefont {A.~W.~C.}\
  \bibnamefont {Lau}}\ and\ \bibinfo {author} {\bibfnamefont {T.~C.}\
  \bibnamefont {Lubensky}},\ }\bibfield  {title} {{\selectlanguage
  {english}\enquote {\bibinfo {title} {State-dependent diffusion: Thermodynamic
  consistency and its path integral formulation},}\ }}\href@noop {} {\bibfield
  {journal} {\bibinfo  {journal} {Phys. Rev. E},\ }\textbf {\bibinfo {volume}
  {76}},\ \bibinfo {pages} {011123} (\bibinfo {year} {2007})}\BibitemShut
  {NoStop}%
\bibitem [{\citenamefont {Horsthemke}\ and\ \citenamefont
  {Lefever}(1984)}]{Lefever:bk:1984}%
  \BibitemOpen
  \bibfield  {author} {\bibinfo {author} {\bibfnamefont {W.}~\bibnamefont
  {Horsthemke}}\ and\ \bibinfo {author} {\bibfnamefont {R.}~\bibnamefont
  {Lefever}},\ }\href@noop {} {{\selectlanguage {english}\emph {\bibinfo
  {title} {Noise-Induced Transitions: Theory and Applications in Physics,
  Chemistry, and Biology}}}},\ Springer Series in Synergetics\ (\bibinfo
  {publisher} {Springer},\ \bibinfo {address} {Berlin},\ \bibinfo {year}
  {1984})\ p.\ \bibinfo {pages} {318}\BibitemShut {NoStop}%
\bibitem [{\citenamefont {van Kampen}(1992)}]{Kampen:bk:1992}%
  \BibitemOpen
  \bibfield  {author} {\bibinfo {author} {\bibfnamefont {N.~G.}\ \bibnamefont
  {van Kampen}},\ }\href@noop {} {{\selectlanguage {english}\emph {\bibinfo
  {title} {Stochastic Processes in Physics and Chemistry}}}},\ \bibinfo
  {edition} {3rd}\ ed.\ (\bibinfo  {publisher} {Elsevier},\ \bibinfo {address}
  {Amsterdam},\ \bibinfo {year} {1992})\ p.\ \bibinfo {pages} {463}\BibitemShut
  {NoStop}%
\bibitem [{\citenamefont {Kurchan}(1998)}]{Kurchan:jpa:1998}%
  \BibitemOpen
  \bibfield  {author} {\bibinfo {author} {\bibfnamefont {J.}~\bibnamefont
  {Kurchan}},\ }\bibfield  {title} {{\selectlanguage {english}\enquote
  {\bibinfo {title} {Fluctuation theorem for stochastic dynamics},}\
  }}\href@noop {} {\bibfield  {journal} {\bibinfo  {journal} {J. Phys. A: Math.
  Gen.},\ }\textbf {\bibinfo {volume} {31}},\ \bibinfo {pages} {3719--3729}
  (\bibinfo {year} {1998})}\BibitemShut {NoStop}%
\bibitem [{\citenamefont {Kurchan}(2007)}]{Kurchan:jsm:2007}%
  \BibitemOpen
  \bibfield  {author} {\bibinfo {author} {\bibfnamefont {J.}~\bibnamefont
  {Kurchan}},\ }\bibfield  {title} {{\selectlanguage {english}\enquote
  {\bibinfo {title} {Non-equilibrium work relations},}\ }}\href@noop {}
  {\bibfield  {journal} {\bibinfo  {journal} {J. Stat. Mech.},\ \bibinfo
  {pages} {P07005}} (\bibinfo {year} {2007})}\BibitemShut {NoStop}%
\bibitem [{\citenamefont {Klimontovich}(1994)}]{Klimont:ufn:1994}%
  \BibitemOpen
  \bibfield  {author} {\bibinfo {author} {\bibfnamefont {Yu.~L.}\ \bibnamefont
  {Klimontovich}},\ }\bibfield  {title} {{\selectlanguage {english}\enquote
  {\bibinfo {title} {Nonlinear {B}rownian motion},}\ }}\href@noop {} {\bibfield
   {journal} {\bibinfo  {journal} {Phys.-Usp.},\ }\textbf {\bibinfo {volume}
  {37}},\ \bibinfo {pages} {737} (\bibinfo {year} {1994})}\BibitemShut
  {NoStop}%
\bibitem [{\citenamefont {Landau}\ and\ \citenamefont
  {Lifshitz}(1987)}]{Landau:6v:en:1987}%
  \BibitemOpen
  \bibfield  {author} {\bibinfo {author} {\bibfnamefont {L.~D.}\ \bibnamefont
  {Landau}}\ and\ \bibinfo {author} {\bibfnamefont {E.~M.}\ \bibnamefont
  {Lifshitz}},\ }\href@noop {} {{\selectlanguage {english}\emph {\bibinfo
  {title} {Fluid Mechanics}}}},\ \bibinfo {edition} {2nd}\ ed.,\ \bibinfo
  {series} {Course of Theoretical Physics}, Vol.~\bibinfo {volume} {6}\
  (\bibinfo  {publisher} {Pergamon Press},\ \bibinfo {address} {NY, USA},\
  \bibinfo {year} {1987})\ p.\ \bibinfo {pages} {539}\BibitemShut {NoStop}%
\bibitem [{\citenamefont {Rayleigh}(1945)}]{Rayleigh:bk:1945}%
  \BibitemOpen
  \bibfield  {author} {\bibinfo {author} {\bibfnamefont {J.~W.~S.}\
  \bibnamefont {Rayleigh}},\ }\href@noop {} {{\selectlanguage {english}\emph
  {\bibinfo {title} {The Theory of Sound}}}},\ \bibinfo {edition} {2nd}\ ed.\
  (\bibinfo  {publisher} {Dover},\ \bibinfo {address} {NY},\ \bibinfo {year}
  {1945})\BibitemShut {NoStop}%
\bibitem [{\citenamefont {Crooks}(1999)}]{Crooks:pre:1999}%
  \BibitemOpen
  \bibfield  {author} {\bibinfo {author} {\bibfnamefont {G.~E.}\ \bibnamefont
  {Crooks}},\ }\bibfield  {title} {{\selectlanguage {english}\enquote {\bibinfo
  {title} {Entropy production fluctuation theorem and the nonequilibrium work
  relation for free energy differences},}\ }}\href@noop {} {\bibfield
  {journal} {\bibinfo  {journal} {Phys. Rev. E},\ }\textbf {\bibinfo {volume}
  {60}},\ \bibinfo {pages} {2721--2726} (\bibinfo {year} {1999})}\BibitemShut
  {NoStop}%
\bibitem [{\citenamefont {Seifert}(2005)}]{Seifert:prl:2005}%
  \BibitemOpen
  \bibfield  {author} {\bibinfo {author} {\bibfnamefont {U.}~\bibnamefont
  {Seifert}},\ }\bibfield  {title} {{\selectlanguage {english}\enquote
  {\bibinfo {title} {Entropy production along a stochastic trajectory and an
  integral fluctuation theorem},}\ }}\href@noop {} {\bibfield  {journal}
  {\bibinfo  {journal} {Phys. Rev. Lett.},\ }\textbf {\bibinfo {volume} {95}},\
  \bibinfo {pages} {040602} (\bibinfo {year} {2005})}\BibitemShut {NoStop}%
\bibitem [{\citenamefont {Vasil'ev}(2004)}]{Vas:bk:2004}%
  \BibitemOpen
  \bibfield  {author} {\bibinfo {author} {\bibfnamefont {A.~N.}\ \bibnamefont
  {Vasil'ev}},\ }\href@noop {} {{\selectlanguage {english}\emph {\bibinfo
  {title} {The Field Theoretic Renormalization Group in Critical Behavior
  Theory and Stochastic Dynamics}}}}\ (\bibinfo  {publisher} {CRC Press},\
  \bibinfo {address} {London},\ \bibinfo {year} {2004})\ p.\ \bibinfo {pages}
  {681}\BibitemShut {NoStop}%
\bibitem [{\citenamefont {Hochberg}\ \emph {et~al.}(1999)\citenamefont
  {Hochberg}, \citenamefont {Molina-Par{\'{\i}}s},\ and\ \citenamefont
  {P{\'e}rez-Mercader}}]{Hoch:pre:1999}%
  \BibitemOpen
  \bibfield  {author} {\bibinfo {author} {\bibfnamefont {D.}~\bibnamefont
  {Hochberg}}, \bibinfo {author} {\bibfnamefont {C.}~\bibnamefont
  {Molina-Par{\'{\i}}s}}, \ and\ \bibinfo {author} {\bibfnamefont
  {J.}~\bibnamefont {P{\'e}rez-Mercader}},\ }\bibfield  {title}
  {{\selectlanguage {english}\enquote {\bibinfo {title} {Effective action for
  stochastic differential equations},}\ }}\href@noop {} {\bibfield  {journal}
  {\bibinfo  {journal} {Phys. Rev. E},\ }\textbf {\bibinfo {volume} {60}},\
  \bibinfo {pages} {6343--6353} (\bibinfo {year} {1999})}\BibitemShut {NoStop}%
\bibitem [{\citenamefont {Kiselev}(2000)}]{Kis:susy}%
  \BibitemOpen
  \bibfield  {author} {\bibinfo {author} {\bibfnamefont {A.~D.}\ \bibnamefont
  {Kiselev}},\ }\bibfield  {title} {{\selectlanguage {english}\enquote
  {\bibinfo {title} {Supersymmetry approach in the field theory of ergodicity
  breaking transitions},}\ }}\href@noop {} {\bibfield  {journal} {\bibinfo
  {journal} {Physica A},\ }\textbf {\bibinfo {volume} {285}},\ \bibinfo {pages}
  {413--432} (\bibinfo {year} {2000})}\BibitemShut {NoStop}%
\bibitem [{\citenamefont {Arnold}(2000)}]{Arnold:pre:2:2000}%
  \BibitemOpen
  \bibfield  {author} {\bibinfo {author} {\bibfnamefont {Peter}\ \bibnamefont
  {Arnold}},\ }\bibfield  {title} {{\selectlanguage {english}\enquote {\bibinfo
  {title} {Symmetric path integrals for stochastic equations with
  multiplicative noise},}\ }}\href@noop {} {\bibfield  {journal} {\bibinfo
  {journal} {Phys. Rev. E},\ }\textbf {\bibinfo {volume} {61}},\ \bibinfo
  {pages} {6099--6102} (\bibinfo {year} {2000})}\BibitemShut {NoStop}%
\bibitem [{\citenamefont {Arenas}\ and\ \citenamefont
  {Barci}(2010)}]{Arenas:pre:2010}%
  \BibitemOpen
  \bibfield  {author} {\bibinfo {author} {\bibfnamefont {Zochil~Gonz\'alez}\
  \bibnamefont {Arenas}}\ and\ \bibinfo {author} {\bibfnamefont {Daniel~G.}\
  \bibnamefont {Barci}},\ }\bibfield  {title} {{\selectlanguage
  {english}\enquote {\bibinfo {title} {Functional integral approach for
  multiplicative stochastic processes},}\ }}\href@noop {} {\bibfield  {journal}
  {\bibinfo  {journal} {Phys. Rev. E},\ }\textbf {\bibinfo {volume} {81}},\
  \bibinfo {pages} {051113} (\bibinfo {year} {2010})}\BibitemShut {NoStop}%
\bibitem [{\citenamefont {Searles}\ and\ \citenamefont
  {Evans}(1999)}]{Searles:pre:1999}%
  \BibitemOpen
  \bibfield  {author} {\bibinfo {author} {\bibfnamefont {D.~J.}\ \bibnamefont
  {Searles}}\ and\ \bibinfo {author} {\bibfnamefont {D.~J.}\ \bibnamefont
  {Evans}},\ }\bibfield  {title} {{\selectlanguage {english}\enquote {\bibinfo
  {title} {Fluctuation theorem for stochastic systems},}\ }}\href@noop {}
  {\bibfield  {journal} {\bibinfo  {journal} {Phys. Rev. E},\ }\textbf
  {\bibinfo {volume} {60}},\ \bibinfo {pages} {159--164} (\bibinfo {year}
  {1999})}\BibitemShut {NoStop}%
\bibitem [{\citenamefont {Evans}\ and\ \citenamefont
  {Searles}(2002)}]{Evans:advph:2002}%
  \BibitemOpen
  \bibfield  {author} {\bibinfo {author} {\bibfnamefont {D.~J.}\ \bibnamefont
  {Evans}}\ and\ \bibinfo {author} {\bibfnamefont {D.~J.}\ \bibnamefont
  {Searles}},\ }\bibfield  {title} {{\selectlanguage {english}\enquote
  {\bibinfo {title} {The fluctuation theorem},}\ }}\href@noop {} {\bibfield
  {journal} {\bibinfo  {journal} {Adv. Phys.},\ }\textbf {\bibinfo {volume}
  {51}},\ \bibinfo {pages} {1529--1585} (\bibinfo {year} {2002})}\BibitemShut
  {NoStop}%
\bibitem [{\citenamefont {Marconi}\ \emph {et~al.}(2008)\citenamefont
  {Marconi}, \citenamefont {Puglisi}, \citenamefont {Rondoni},\ and\
  \citenamefont {Vulpiani}}]{Marconi:phrep:2008}%
  \BibitemOpen
  \bibfield  {author} {\bibinfo {author} {\bibfnamefont {U.~M.~B.}\
  \bibnamefont {Marconi}}, \bibinfo {author} {\bibfnamefont {A.}~\bibnamefont
  {Puglisi}}, \bibinfo {author} {\bibfnamefont {L.}~\bibnamefont {Rondoni}}, \
  and\ \bibinfo {author} {\bibfnamefont {A.}~\bibnamefont {Vulpiani}},\
  }\bibfield  {title} {{\selectlanguage {english}\enquote {\bibinfo {title}
  {Fluctuation-dissipation: {R}esponse theory in statistical physics},}\
  }}\href@noop {} {\bibfield  {journal} {\bibinfo  {journal} {Phys. Rep.},\
  }\textbf {\bibinfo {volume} {461}},\ \bibinfo {pages} {111--195} (\bibinfo
  {year} {2008})}\BibitemShut {NoStop}%
\bibitem [{\citenamefont {Esposito}\ \emph {et~al.}(2009)\citenamefont
  {Esposito}, \citenamefont {Harbola},\ and\ \citenamefont
  {Mukamel}}]{Esposito:rmp:2009}%
  \BibitemOpen
  \bibfield  {author} {\bibinfo {author} {\bibfnamefont {M.}~\bibnamefont
  {Esposito}}, \bibinfo {author} {\bibfnamefont {U.}~\bibnamefont {Harbola}}, \
  and\ \bibinfo {author} {\bibfnamefont {S.}~\bibnamefont {Mukamel}},\
  }\bibfield  {title} {{\selectlanguage {english}\enquote {\bibinfo {title}
  {Nonequilibrium fluctuations, fluctuation theorems, counting statistics in
  quantum systems},}\ }}\href@noop {} {\bibfield  {journal} {\bibinfo
  {journal} {Rev. Mod. Phys.},\ }\textbf {\bibinfo {volume} {81}},\ \bibinfo
  {pages} {1665--1702} (\bibinfo {year} {2009})}\BibitemShut {NoStop}%
\bibitem [{\citenamefont {Wang}\ \emph {et~al.}(2002)\citenamefont {Wang},
  \citenamefont {Sevick}, \citenamefont {Mittag}, \citenamefont {Searles},\
  and\ \citenamefont {Evans}}]{Wang:prl:2002}%
  \BibitemOpen
  \bibfield  {author} {\bibinfo {author} {\bibfnamefont {G.~M.}\ \bibnamefont
  {Wang}}, \bibinfo {author} {\bibfnamefont {E.~M.}\ \bibnamefont {Sevick}},
  \bibinfo {author} {\bibfnamefont {Emil}\ \bibnamefont {Mittag}}, \bibinfo
  {author} {\bibfnamefont {Debra~J.}\ \bibnamefont {Searles}}, \ and\ \bibinfo
  {author} {\bibfnamefont {Denis~J.}\ \bibnamefont {Evans}},\ }\bibfield
  {title} {{\selectlanguage {english}\enquote {\bibinfo {title} {Experimental
  demonstration of violations of the second law of thermodynamics for small
  systems and short time scales},}\ }}\href@noop {} {\bibfield  {journal}
  {\bibinfo  {journal} {Phys. Rev. Lett.},\ }\textbf {\bibinfo {volume} {89}},\
  \bibinfo {pages} {050601} (\bibinfo {year} {2002})}\BibitemShut {NoStop}%
\bibitem [{\citenamefont {Carberry}\ \emph {et~al.}(2004)\citenamefont
  {Carberry}, \citenamefont {Reid}, \citenamefont {Wang}, \citenamefont
  {Sevick}, \citenamefont {Searles},\ and\ \citenamefont
  {Evans}}]{Carberry:prl:2004}%
  \BibitemOpen
  \bibfield  {author} {\bibinfo {author} {\bibfnamefont {D.~M.}\ \bibnamefont
  {Carberry}}, \bibinfo {author} {\bibfnamefont {J.~C.}\ \bibnamefont {Reid}},
  \bibinfo {author} {\bibfnamefont {G.~M.}\ \bibnamefont {Wang}}, \bibinfo
  {author} {\bibfnamefont {E.~M.}\ \bibnamefont {Sevick}}, \bibinfo {author}
  {\bibfnamefont {D.~J.}\ \bibnamefont {Searles}}, \ and\ \bibinfo {author}
  {\bibfnamefont {D.~J.}\ \bibnamefont {Evans}},\ }\bibfield  {title}
  {{\selectlanguage {english}\enquote {\bibinfo {title} {Fluctuations and
  irreversibility: an experimental demonstration of a second-law-like theorem
  using a colloidal particle held in an optical trap},}\ }}\href@noop {}
  {\bibfield  {journal} {\bibinfo  {journal} {Phys. Rev. Lett.},\ }\textbf
  {\bibinfo {volume} {92}},\ \bibinfo {pages} {140601} (\bibinfo {year}
  {2004})}\BibitemShut {NoStop}%
\bibitem [{\citenamefont {Trepagnier}\ \emph {et~al.}(2004)\citenamefont
  {Trepagnier}, \citenamefont {Jarzynski}, \citenamefont {Ritort},
  \citenamefont {Crooks}, \citenamefont {Bustamante},\ and\ \citenamefont
  {Liphardt}}]{Trepag:pnas:2004}%
  \BibitemOpen
  \bibfield  {author} {\bibinfo {author} {\bibfnamefont {E.~H.}\ \bibnamefont
  {Trepagnier}}, \bibinfo {author} {\bibfnamefont {C.}~\bibnamefont
  {Jarzynski}}, \bibinfo {author} {\bibfnamefont {F.}~\bibnamefont {Ritort}},
  \bibinfo {author} {\bibfnamefont {G.~E.}\ \bibnamefont {Crooks}}, \bibinfo
  {author} {\bibfnamefont {C.~J.}\ \bibnamefont {Bustamante}}, \ and\ \bibinfo
  {author} {\bibfnamefont {J.}~\bibnamefont {Liphardt}},\ }\bibfield  {title}
  {{\selectlanguage {english}\enquote {\bibinfo {title} {Experimental test of
  {H}atano and {S}asa's nonequilibrium steady-state equality},}\ }}\href@noop
  {} {\bibfield  {journal} {\bibinfo  {journal} {Proc. Natl. Acad. Sci. USA},\
  }\textbf {\bibinfo {volume} {101}},\ \bibinfo {pages} {15038--15041}
  (\bibinfo {year} {2004})}\BibitemShut {NoStop}%
\bibitem [{\citenamefont {Wang}\ \emph {et~al.}(2005)\citenamefont {Wang},
  \citenamefont {Reid}, \citenamefont {Carberry}, \citenamefont {Williams},
  \citenamefont {Sevick},\ and\ \citenamefont {Evans}}]{Wang:pre:2005}%
  \BibitemOpen
  \bibfield  {author} {\bibinfo {author} {\bibfnamefont {G.~M.}\ \bibnamefont
  {Wang}}, \bibinfo {author} {\bibfnamefont {J.~C.}\ \bibnamefont {Reid}},
  \bibinfo {author} {\bibfnamefont {D.~M.}\ \bibnamefont {Carberry}}, \bibinfo
  {author} {\bibfnamefont {D.~R.~M.}\ \bibnamefont {Williams}}, \bibinfo
  {author} {\bibfnamefont {E.~M.}\ \bibnamefont {Sevick}}, \ and\ \bibinfo
  {author} {\bibfnamefont {Denis~J.}\ \bibnamefont {Evans}},\ }\bibfield
  {title} {{\selectlanguage {english}\enquote {\bibinfo {title} {Experimental
  study of the fluctuation theorem in a nonequilibrium steady state},}\
  }}\href@noop {} {\bibfield  {journal} {\bibinfo  {journal} {Phys. Rev. E},\
  }\textbf {\bibinfo {volume} {71}},\ \bibinfo {pages} {046142} (\bibinfo
  {year} {2005})}\BibitemShut {NoStop}%
\bibitem [{\citenamefont {Blickle}\ \emph {et~al.}(2006)\citenamefont
  {Blickle}, \citenamefont {Speck}, \citenamefont {Helden}, \citenamefont
  {Seifert},\ and\ \citenamefont {Bechinger}}]{Blickle:prl:2006}%
  \BibitemOpen
  \bibfield  {author} {\bibinfo {author} {\bibfnamefont {V.}~\bibnamefont
  {Blickle}}, \bibinfo {author} {\bibfnamefont {T.}~\bibnamefont {Speck}},
  \bibinfo {author} {\bibfnamefont {L.}~\bibnamefont {Helden}}, \bibinfo
  {author} {\bibfnamefont {U.}~\bibnamefont {Seifert}}, \ and\ \bibinfo
  {author} {\bibfnamefont {C.}~\bibnamefont {Bechinger}},\ }\bibfield  {title}
  {{\selectlanguage {english}\enquote {\bibinfo {title} {Thermodynamics of a
  colloidal particle in a time-dependent nonharmonic potential},}\ }}\href@noop
  {} {\bibfield  {journal} {\bibinfo  {journal} {Phys. Rev. Lett.},\ }\textbf
  {\bibinfo {volume} {96}},\ \bibinfo {pages} {070603} (\bibinfo {year}
  {2006})}\BibitemShut {NoStop}%
\bibitem [{\citenamefont {Carberry}\ \emph {et~al.}(2007)\citenamefont
  {Carberry}, \citenamefont {Baker}, \citenamefont {Wang}, \citenamefont
  {Sevick},\ and\ \citenamefont {Evans}}]{Carberry:joa:2007}%
  \BibitemOpen
  \bibfield  {author} {\bibinfo {author} {\bibfnamefont {D.~M.}\ \bibnamefont
  {Carberry}}, \bibinfo {author} {\bibfnamefont {M.~A.~B.}\ \bibnamefont
  {Baker}}, \bibinfo {author} {\bibfnamefont {G.~M.}\ \bibnamefont {Wang}},
  \bibinfo {author} {\bibfnamefont {E.~M.}\ \bibnamefont {Sevick}}, \ and\
  \bibinfo {author} {\bibfnamefont {D.~J.}\ \bibnamefont {Evans}},\ }\bibfield
  {title} {{\selectlanguage {english}\enquote {\bibinfo {title} {An optical
  trap experiment to demonstrate fluctuation theorems in viscoelastic media},}\
  }}\href@noop {} {\bibfield  {journal} {\bibinfo  {journal} {J. Opt. A: Pure
  Appl. Opt.},\ }\textbf {\bibinfo {volume} {9}},\ \bibinfo {pages}
  {S204--S214} (\bibinfo {year} {2007})}\BibitemShut {NoStop}%
\bibitem [{\citenamefont {Andrieux}\ \emph {et~al.}(2007)\citenamefont
  {Andrieux}, \citenamefont {Gaspard}, \citenamefont {Ciliberto}, \citenamefont
  {Garnier}, \citenamefont {Joubaud},\ and\ \citenamefont
  {Petrosyan}}]{Andrieux:prl:2007}%
  \BibitemOpen
  \bibfield  {author} {\bibinfo {author} {\bibfnamefont {D.}~\bibnamefont
  {Andrieux}}, \bibinfo {author} {\bibfnamefont {P.}~\bibnamefont {Gaspard}},
  \bibinfo {author} {\bibfnamefont {S.}~\bibnamefont {Ciliberto}}, \bibinfo
  {author} {\bibfnamefont {N.}~\bibnamefont {Garnier}}, \bibinfo {author}
  {\bibfnamefont {S.}~\bibnamefont {Joubaud}}, \ and\ \bibinfo {author}
  {\bibfnamefont {A.}~\bibnamefont {Petrosyan}},\ }\bibfield  {title}
  {{\selectlanguage {english}\enquote {\bibinfo {title} {Entropy production and
  time asymmetry in nonequilibrium fluctuations},}\ }}\href@noop {} {\bibfield
  {journal} {\bibinfo  {journal} {Phys. Rev. Lett.},\ }\textbf {\bibinfo
  {volume} {98}},\ \bibinfo {pages} {150601} (\bibinfo {year}
  {2007})}\BibitemShut {NoStop}%
\bibitem [{\citenamefont {Liphardt}\ \emph {et~al.}(2002)\citenamefont
  {Liphardt}, \citenamefont {Dumont}, \citenamefont {Smith}, \citenamefont
  {Jr.},\ and\ \citenamefont {Bustamante}}]{Liphardt:sci:2002}%
  \BibitemOpen
  \bibfield  {author} {\bibinfo {author} {\bibfnamefont {J.}~\bibnamefont
  {Liphardt}}, \bibinfo {author} {\bibfnamefont {S.}~\bibnamefont {Dumont}},
  \bibinfo {author} {\bibfnamefont {S.~B.}\ \bibnamefont {Smith}}, \bibinfo
  {author} {\bibfnamefont {I.~Tinoco}\ \bibnamefont {Jr.}}, \ and\ \bibinfo
  {author} {\bibfnamefont {C.}~\bibnamefont {Bustamante}},\ }\bibfield  {title}
  {{\selectlanguage {english}\enquote {\bibinfo {title} {Equilibrium
  information from nonequilibrium measurements in an experimental test of
  {J}arzynski's equality},}\ }}\href@noop {} {\bibfield  {journal} {\bibinfo
  {journal} {Science},\ }\textbf {\bibinfo {volume} {296}},\ \bibinfo {pages}
  {1832--1835} (\bibinfo {year} {2002})}\BibitemShut {NoStop}%
\bibitem [{\citenamefont {Collin}\ \emph {et~al.}(2005)\citenamefont {Collin},
  \citenamefont {Ritort}, \citenamefont {Jarzynski}, \citenamefont {Smith},
  \citenamefont {Jr.},\ and\ \citenamefont {Bustamante}}]{Collin:nat:2005}%
  \BibitemOpen
  \bibfield  {author} {\bibinfo {author} {\bibfnamefont {D.}~\bibnamefont
  {Collin}}, \bibinfo {author} {\bibfnamefont {F.}~\bibnamefont {Ritort}},
  \bibinfo {author} {\bibfnamefont {C.}~\bibnamefont {Jarzynski}}, \bibinfo
  {author} {\bibfnamefont {S.~B.}\ \bibnamefont {Smith}}, \bibinfo {author}
  {\bibfnamefont {I.~Tinoco}\ \bibnamefont {Jr.}}, \ and\ \bibinfo {author}
  {\bibfnamefont {C.}~\bibnamefont {Bustamante}},\ }\bibfield  {title}
  {{\selectlanguage {english}\enquote {\bibinfo {title} {Verification of the
  {C}rooks fluctuation theorem and recovery of {RNA} folding free energy},}\
  }}\href@noop {} {\bibfield  {journal} {\bibinfo  {journal} {Nature},\
  }\textbf {\bibinfo {volume} {437}},\ \bibinfo {pages} {231--234} (\bibinfo
  {year} {2005})}\BibitemShut {NoStop}%
\bibitem [{\citenamefont {Gomez-Solano}\ \emph {et~al.}(2010)\citenamefont
  {Gomez-Solano}, \citenamefont {Bellon}, \citenamefont {Petrosyan},\ and\
  \citenamefont {Ciliberto}}]{Solano:epl:2010}%
  \BibitemOpen
  \bibfield  {author} {\bibinfo {author} {\bibfnamefont {J.~R.}\ \bibnamefont
  {Gomez-Solano}}, \bibinfo {author} {\bibfnamefont {L.}~\bibnamefont
  {Bellon}}, \bibinfo {author} {\bibfnamefont {A.}~\bibnamefont {Petrosyan}}, \
  and\ \bibinfo {author} {\bibfnamefont {S.}~\bibnamefont {Ciliberto}},\
  }\bibfield  {title} {{\selectlanguage {english}\enquote {\bibinfo {title}
  {Steady-state fluctuation relations for systems driven by an external random
  force},}\ }}\href@noop {} {\bibfield  {journal} {\bibinfo  {journal}
  {Europhys. Lett.},\ }\textbf {\bibinfo {volume} {89}},\ \bibinfo {pages}
  {60003} (\bibinfo {year} {2010})}\BibitemShut {NoStop}%
\bibitem [{\citenamefont {Hatano}\ and\ \citenamefont
  {Sasa}(2001)}]{Hatano:prl:2001}%
  \BibitemOpen
  \bibfield  {author} {\bibinfo {author} {\bibfnamefont {T.}~\bibnamefont
  {Hatano}}\ and\ \bibinfo {author} {\bibfnamefont {S.-I.}\ \bibnamefont
  {Sasa}},\ }\bibfield  {title} {{\selectlanguage {english}\enquote {\bibinfo
  {title} {Steady-state thermodynamics of {L}angevin systems},}\ }}\href@noop
  {} {\bibfield  {journal} {\bibinfo  {journal} {Phys. Rev. Lett.},\ }\textbf
  {\bibinfo {volume} {86}},\ \bibinfo {pages} {3463--3466} (\bibinfo {year}
  {2001})}\BibitemShut {NoStop}%
\bibitem [{\citenamefont {Jarzynski}(1997)}]{Jarzyn:prl:1997}%
  \BibitemOpen
  \bibfield  {author} {\bibinfo {author} {\bibfnamefont {C.}~\bibnamefont
  {Jarzynski}},\ }\bibfield  {title} {{\selectlanguage {english}\enquote
  {\bibinfo {title} {Nonequilibrium equality for free energy differences},}\
  }}\href@noop {} {\bibfield  {journal} {\bibinfo  {journal} {Phys. Rev.
  Lett.},\ }\textbf {\bibinfo {volume} {78}},\ \bibinfo {pages} {2690--2693}
  (\bibinfo {year} {1997})}\BibitemShut {NoStop}%
\bibitem [{\citenamefont {Lebowitz}\ and\ \citenamefont
  {Spohn}(1999)}]{Lebowitz:jsp:1999}%
  \BibitemOpen
  \bibfield  {author} {\bibinfo {author} {\bibfnamefont {J.~L.}\ \bibnamefont
  {Lebowitz}}\ and\ \bibinfo {author} {\bibfnamefont {H.}~\bibnamefont
  {Spohn}},\ }\bibfield  {title} {{\selectlanguage {english}\enquote {\bibinfo
  {title} {A {G}allavotti-{C}ohen-type symmetry in large deviation functional
  for stochastic dynamics},}\ }}\href@noop {} {\bibfield  {journal} {\bibinfo
  {journal} {J. Stat. Phys.},\ }\textbf {\bibinfo {volume} {95}},\ \bibinfo
  {pages} {333--365} (\bibinfo {year} {1999})}\BibitemShut {NoStop}%
\bibitem [{\citenamefont {Maes}(1999)}]{Maes:jsp:1999}%
  \BibitemOpen
  \bibfield  {author} {\bibinfo {author} {\bibfnamefont {C.}~\bibnamefont
  {Maes}},\ }\bibfield  {title} {{\selectlanguage {english}\enquote {\bibinfo
  {title} {The fluctuation theorem as a {G}ibbs property},}\ }}\href@noop {}
  {\bibfield  {journal} {\bibinfo  {journal} {J. Stat. Phys.},\ }\textbf
  {\bibinfo {volume} {95}},\ \bibinfo {pages} {367--392} (\bibinfo {year}
  {1999})}\BibitemShut {NoStop}%
\bibitem [{\citenamefont {Ge}\ and\ \citenamefont {Jiang}(2007)}]{Ge:jpa:2007}%
  \BibitemOpen
  \bibfield  {author} {\bibinfo {author} {\bibfnamefont {Hao}\ \bibnamefont
  {Ge}}\ and\ \bibinfo {author} {\bibfnamefont {Da-Quan}\ \bibnamefont
  {Jiang}},\ }\bibfield  {title} {{\selectlanguage {english}\enquote {\bibinfo
  {title} {The transient fluctuation theorem of sample entropy production for
  general stochastic processes},}\ }}\href@noop {} {\bibfield  {journal}
  {\bibinfo  {journal} {J. Phys. A: Math. Theor.},\ }\textbf {\bibinfo {volume}
  {40}},\ \bibinfo {pages} {F713--F723} (\bibinfo {year} {2007})}\BibitemShut
  {NoStop}%
\bibitem [{\citenamefont {Ge}\ and\ \citenamefont {Jiang}(2008)}]{Ge:jsp:2008}%
  \BibitemOpen
  \bibfield  {author} {\bibinfo {author} {\bibfnamefont {Hao}\ \bibnamefont
  {Ge}}\ and\ \bibinfo {author} {\bibfnamefont {Da-Quan}\ \bibnamefont
  {Jiang}},\ }\bibfield  {title} {{\selectlanguage {english}\enquote {\bibinfo
  {title} {Generalized {J}arzynski's equality of inhomogeneous multidimensional
  diffusion process},}\ }}\href@noop {} {\bibfield  {journal} {\bibinfo
  {journal} {J. Stat. Phys.},\ }\textbf {\bibinfo {volume} {131}},\ \bibinfo
  {pages} {675--689} (\bibinfo {year} {2008})}\BibitemShut {NoStop}%
\bibitem [{\citenamefont {Shargel}(2010)}]{Shargel:jpa:2010}%
  \BibitemOpen
  \bibfield  {author} {\bibinfo {author} {\bibfnamefont {Benjamin~Hertz}\
  \bibnamefont {Shargel}},\ }\bibfield  {title} {{\selectlanguage
  {english}\enquote {\bibinfo {title} {The measure-theoretic identity
  underlying transient fluctuation theorems},}\ }}\href@noop {} {\bibfield
  {journal} {\bibinfo  {journal} {J. Phys. A: Math. Theor.},\ }\textbf
  {\bibinfo {volume} {43}},\ \bibinfo {pages} {135002} (\bibinfo {year}
  {2010})}\BibitemShut {NoStop}%
\bibitem [{\citenamefont {Speck}\ and\ \citenamefont
  {Seifert}(2007)}]{Speck:jsm:2007}%
  \BibitemOpen
  \bibfield  {author} {\bibinfo {author} {\bibfnamefont {T.}~\bibnamefont
  {Speck}}\ and\ \bibinfo {author} {\bibfnamefont {U.}~\bibnamefont
  {Seifert}},\ }\bibfield  {title} {{\selectlanguage {english}\enquote
  {\bibinfo {title} {The {J}arzynski relation, fluctuation theorems, and
  stochastic thermodynamics for non-{M}arkovian processes},}\ }}\href@noop {}
  {\bibfield  {journal} {\bibinfo  {journal} {J. Stat. Mech.},\ \bibinfo
  {pages} {L09002}} (\bibinfo {year} {2007})}\BibitemShut {NoStop}%
\bibitem [{\citenamefont {Seifert}(2008)}]{Seifert:epjb:2008}%
  \BibitemOpen
  \bibfield  {author} {\bibinfo {author} {\bibfnamefont {U.}~\bibnamefont
  {Seifert}},\ }\bibfield  {title} {{\selectlanguage {english}\enquote
  {\bibinfo {title} {Stochastic thermodynamics: {P}rinciples and
  perspectives},}\ }}\href@noop {} {\bibfield  {journal} {\bibinfo  {journal}
  {Eur. Phys. J. B},\ }\textbf {\bibinfo {volume} {64}},\ \bibinfo {pages}
  {423--431} (\bibinfo {year} {2008})}\BibitemShut {NoStop}%
\bibitem [{\citenamefont {Prost}\ \emph {et~al.}(2009)\citenamefont {Prost},
  \citenamefont {Joanny},\ and\ \citenamefont {Parrondo}}]{Prost:prl:2009}%
  \BibitemOpen
  \bibfield  {author} {\bibinfo {author} {\bibfnamefont {J.}~\bibnamefont
  {Prost}}, \bibinfo {author} {\bibfnamefont {J.-F.}\ \bibnamefont {Joanny}}, \
  and\ \bibinfo {author} {\bibfnamefont {J.~M.~R.}\ \bibnamefont {Parrondo}},\
  }\bibfield  {title} {{\selectlanguage {english}\enquote {\bibinfo {title}
  {Generalized fluctuation-dissipation theorem for steady-state systems},}\
  }}\href@noop {} {\bibfield  {journal} {\bibinfo  {journal} {Phys. Rev.
  Lett.},\ }\textbf {\bibinfo {volume} {103}},\ \bibinfo {pages} {090601}
  (\bibinfo {year} {2009})}\BibitemShut {NoStop}%
\bibitem [{\citenamefont {Lev}\ and\ \citenamefont
  {Zagorodny}(2009)}]{Lev:pla:2009}%
  \BibitemOpen
  \bibfield  {author} {\bibinfo {author} {\bibfnamefont {B.~I.}\ \bibnamefont
  {Lev}}\ and\ \bibinfo {author} {\bibfnamefont {A.~G.}\ \bibnamefont
  {Zagorodny}},\ }\bibfield  {title} {{\selectlanguage {english}\enquote
  {\bibinfo {title} {Structure formation in system of brownian particle in
  dusty plasma},}\ }}\href@noop {} {\bibfield  {journal} {\bibinfo  {journal}
  {Phys. Lett. A},\ }\textbf {\bibinfo {volume} {373}},\ \bibinfo {pages}
  {1101--1104} (\bibinfo {year} {2009})}\BibitemShut {NoStop}%
\bibitem [{\citenamefont {Chernyak}\ \emph {et~al.}(2006)\citenamefont
  {Chernyak}, \citenamefont {Chertkov},\ and\ \citenamefont
  {Jarzynski}}]{Chernyak:jsm:2006}%
  \BibitemOpen
  \bibfield  {author} {\bibinfo {author} {\bibfnamefont {V.~Y.}\ \bibnamefont
  {Chernyak}}, \bibinfo {author} {\bibfnamefont {M.}~\bibnamefont {Chertkov}},
  \ and\ \bibinfo {author} {\bibfnamefont {C.}~\bibnamefont {Jarzynski}},\
  }\bibfield  {title} {{\selectlanguage {english}\enquote {\bibinfo {title}
  {Path-integral analysis of fluctuation theorems for general {L}angevin
  processes},}\ }}\href@noop {} {\bibfield  {journal} {\bibinfo  {journal} {J.
  Stat. Mech.},\ \bibinfo {pages} {P08001}} (\bibinfo {year}
  {2006})}\BibitemShut {NoStop}%
\bibitem [{\citenamefont {Gomez-Marin}\ \emph {et~al.}(2008)\citenamefont
  {Gomez-Marin}, \citenamefont {Parrondo},\ and\ \citenamefont {den
  Broeck}}]{Marin:epl:2008}%
  \BibitemOpen
  \bibfield  {author} {\bibinfo {author} {\bibfnamefont {A.}~\bibnamefont
  {Gomez-Marin}}, \bibinfo {author} {\bibfnamefont {J.~M.~R.}\ \bibnamefont
  {Parrondo}}, \ and\ \bibinfo {author} {\bibfnamefont {C.~Van}\ \bibnamefont
  {den Broeck}},\ }\bibfield  {title} {{\selectlanguage {english}\enquote
  {\bibinfo {title} {The "footprint" of irreversibility},}\ }}\href@noop {}
  {\bibfield  {journal} {\bibinfo  {journal} {Europhys. Lett.},\ }\textbf
  {\bibinfo {volume} {82}},\ \bibinfo {pages} {50003} (\bibinfo {year}
  {2008})}\BibitemShut {NoStop}%
\bibitem [{\citenamefont {Esposito}\ and\ \citenamefont {den
  Broeck}(2010)}]{Esposito:prl:2010}%
  \BibitemOpen
  \bibfield  {author} {\bibinfo {author} {\bibfnamefont {Massimiliano}\
  \bibnamefont {Esposito}}\ and\ \bibinfo {author} {\bibfnamefont
  {Christian~Van}\ \bibnamefont {den Broeck}},\ }\bibfield  {title}
  {{\selectlanguage {english}\enquote {\bibinfo {title} {Three detailed
  fluctuation theorems},}\ }}\href@noop {} {\bibfield  {journal} {\bibinfo
  {journal} {Phys. Rev. Lett.},\ }\textbf {\bibinfo {volume} {104}},\ \bibinfo
  {pages} {090601} (\bibinfo {year} {2010})}\BibitemShut {NoStop}%
\end{thebibliography}

%merlin.mbs 2010-03-15 4.21a (PWD, AO, DPC)
%Control: key (0)
%Control: author (0) dotless jnrlst
%Control: editor formatted (1) identically to author
%Control: production of article title (0) allowed
%Control: page (1) range
%Control: year (0) verbatim
%Control: production of eprint (0) enabled
%

\end{document}